\documentclass[fleqn,usenatbib,dvipsnames]{mnras}

\usepackage{newtxtext,newtxmath}
\usepackage[T1]{fontenc}

\DeclareRobustCommand{\VAN}[3]{#2}
\let\VANthebibliography\thebibliography
\def\thebibliography{\DeclareRobustCommand{\VAN}[3]{##3}\VANthebibliography}

\usepackage{graphicx}	
\usepackage{amsmath}	
\usepackage[usenames, dvipsnames]{color}

\def\msun{M$_{\odot}$}

\def\mbh{$M\mathrm{_{BH}}$}

\def\LLEdd{$L\mathrm{_{bol}}/L\mathrm{_{Edd}}$}

\title[Black hole mass scaled by coronal gas]{A novel black-hole mass scaling relation based  on coronal gas, and its dependence  with the  accretion disc }

\author[M. A. Prieto et al.]{
M. Almudena Prieto,$^{1}$\thanks{E-mail: aprieto@iac.es}
Alberto Rodr\'{\i}guez-Ardila,$^{2,3}$
Swayamtrupta Panda,$^{2,4,5}$ 
\newauthor Murilo Marinello$^{2}$
\\
$^{1}$ Instituto de Astrof\'{\i}sica de Canarias \& Universidad de La Laguna, E-38205, Tenerife, Spain\\
$^{2}$ LNA/MCTIC, Rua dos Estados Unidos, 154. Bairro das Na\c c\~oes, Itajub\'a, MG 37501-591, Brazil\\
$^{3}$ Divis\~ao de Astrof\'{\i}sica, INPE, Avenida dos Astronautas 1758, S\~ao Jos\'e dos Campos, 12227-010, SP, Brazil \\
$^{4}$ Center for Theoretical Physics, Polish Academy of Sciences, Al. Lotnik{\'o}w 32/46, 02-668 Warsaw, Poland \\
$^{5}$ Nicolaus Copernicus Astronomical Center, Polish Academy of Sciences, ul. Bartycka 18, 00-716 Warsaw, Poland
}

\date{Accepted XXX. Received YYY; in original form ZZZ}

\pubyear{2021}

\begin{document}
\label{firstpage}
\pagerange{\pageref{firstpage}--\pageref{lastpage}}
\maketitle

\begin{abstract}
Using \textit{bona-fide} black hole (BH) mass estimates from reverberation mapping and the line ratio [\ion{Si}{vi}]~1.963$\micron$/Br$\gamma_{\rm broad}$ as  tracer of the AGN ionising continuum, a novel BH-mass scaling relation of the form log($M_{\rm BH}) = (6.40\pm 0.17) - (1.99\pm 0.37) \times$ log ([\ion{Si}{vi}]/Br$\gamma_{\rm broad})$,  dispersion 0.47~dex, over the BH mass interval, $10^6 - 10^8$ M$_{\odot}$  is found. Following on the geometrically thin accretion disc approximation and after surveying a basic parameter space for coronal lines production, we believe one of  main drivers of the  relation is the  effective temperature of the  disc, which is effectively sampled by the [\ion{Si}{vi}]~1.963$\micron$ coronal line for the range of BH masses considered. By means of {\sc CLOUDY} photoionisation models, the observed anti-correlation appears to be formally in line with the thin disc  prediction $T_{\rm disc} \propto {M_{\rm BH}}^{-1/4}$.
\end{abstract}



\section{Introduction} \label{sec:intro}

The determination of black hole (BH) masses is a major focus in the community. Most accurate determinations rely on dynamical analysis, for galaxies including our Milky-Way. That approach, though, gets restricted to relative nearby objects, where high angular observations resolving the BH radius of influence or extensive variability monitoring programs of the broad line region in active galactic nuclei (AGN) are possible \citep[e.g.][]{peterson1993,genzel_etal2010,kormendy_ho2013,bentz+15}. Most BH mass estimates are based however on powerful correlations between the BH mass and  the bulge stellar velocity  dispersion - the M-$\sigma$ relation \citep[e.g.][]{ferrarese_merritt2000,gultekin+09}, or the AGN continuum luminosity - so called mass-luminosity relation  by which the optical, UV or X-ray luminosity are found to correlate with the Broad-Line-Region (BLR) size \citep[e.g.][and references therein]{koratkar_gaskell1991,kaspi_etal2000,kaspi_etal2005,landt+13}.

While the use of the M-$\sigma$ relation requires the measurement of $\sigma$, it is not always easy to determine it, particularly in AGN. In these objects, the strong continuum from the nuclear region dilutes the stellar absorption lines. In order to overcome this difficulty, a number
of alternative scaling relations using emission lines such as [\ion{O}{iii}]~$\lambda$5007 to measure the mass of the bulge 
\citep[e.g.][]{nelson_whittle1996},
 [\ion{O}{ii}]~$\lambda$3727 \citep[e.g.][]{salviander_etal2006}, H$\beta$ or H$\alpha$ \citep[e.g.][]{kaspi_etal2005,greene_ho2005} to infer on the BLR size, or [\ion{Fe}{ii}] in the near-infrared  \citep[e.g.][]{riffel_etal2013} to infer on the stellar $\sigma$,  have been proposed. 
 
Overall, the scatter of the scaling relations  in the literature is   40\% or larger. A  fraction of the scatter should be intrinsic, inherent  to the nature of the parameters used in the relations and their variations between objects.  The Hubble  type and chiefly, the  presence of    bulges or pseudo bulges affect  the M-$\sigma$ relation \citep{gultekin+09}. An intrinsic scatter of up 40\% in the   continuum luminosity - BLR size scaling relation is inferred by \citet{kaspi_etal2005}, mostly introduced  by changes in the  optical - UV continuum shape with increasing AGN luminosity. Other factors such as intrinsic differences in the BLR density and the ionisation parameter are also expected to play a role \citep{collin1988b,Marziani2019,panda_cafe2}.

With the start of operations of  facilities optimised for the near - mid Infrared (IR)  region,  the James-Webb Space Telescope \citep[e.g.][]{JWST2006} and  the Vera Rubin Observatory's LSST \citep[e.g.][]{ivezic_etal2019}, the use of scaling-relations focused on spectral features centred in that interval will be an asset. This paper presents a novel approach to estimate BH masses as a function of high ionization IR emission lines, namely the   coronal lines, after   normalisation to HI broad line emission, so far for Type~I AGN only. The motivation behind relies on the high ionisation potential, IP, of coronal lines, above 50~eV up to few hundreds eV which makes them    excellent tracers of the ionising continuum. Hence,  their ability to sample the peak - or highest temperature, of the ionising continuum whether  this is dominated by a thermal component, an accretion disc or black body modified spectrum.

Coronal lines (CL) spread over the X-rays, optical and IR spectrum. Although often fainter than the classical medium-ionisation lines used for photoionisation diagnosis, high angular resolution in nearby AGN has shown that CL  particularly in the near-IR  are among the most conspicuous features \citep[e.g.][]{marconi1994, reunanen03, prieto+05, rodriguez-ardila06, muller-sanchez+11, rodriguez+17, gravity2020}. 

In this work, we explore possible dependencies of the BH mass with optical and infrared coronal emission  of different  IP after normalising of the coronal emission  to the nearest H\,{\sc i} broad line emission.  
The CLs employed are the most common and brightest ones   in AGN \citep{reunanen03, rodriguez+11, lamperti+17}.
A tight correlation between BH mass and the CL ratio [Si\,{\sc vi}]~1.96~$\mu$m/~Br${\gamma}$  is observed. Possible explanations for the observed correlation ( and the  absence  of it for some  other CLs)  are examined in the context of accretion theory and photoionisation model predictions.

The paper is organised as follows. Sect.~\ref{sec:obs} describes the observations and data reduction employed in this work; Sect.~\ref{sec3} deals with the diagnostics diagrams developed by us to weigh the BH mass using CL; Sect.~\ref{accretion-disc} examines the scaling relationship in the framework of accretion theory and photoionisation predictions.  The implications and limitations of the results    are presented in Sect.~\ref{sec:remarks}. Throughout this
 paper, the following cosmology is adopted: H$_{\rm o}=70$~km\,s$^{-1}$~Mpc$^{-1}$; $\Omega_M$=0.30; $\Omega_{\Lambda}=0.70$.

\section{Data selection}
\label{sec:obs}
Objects in this work are selected by having BH masses determined by reverberation mapping and single epoch optical and/or near-IR spectra with accurate CL measurements.  The first criterion restricts the sample to Type~I sources only. The second avoids variability issues. Although we give preference to sources with both optical and near-IR spectra available, this final criterion could not always be fulfilled.  

The CL used are [\ion{Fe}{vii}]~$\lambda$6087~\AA\ in the optical and [S\,{\sc viii}]~0.991~$\mu$m, [Si\,{\sc x}]~1.432~$\mu$m and [Si\,{\sc vi}]~1.964~$\mu$m in the near-IR. They are among the strongest CL in AGN \citep{reunanen03, rodriguez+11, lamperti+17} and span a wide IP range, 100 - 350 ~eV. In addition,  H\,{\sc ii} lines of H$\beta$, Pa$\beta$ and Br$\gamma$ are employed. The whole set samples the ionising continuum over the 13.6 - 351~eV range.  Near-IR CL were preferred because of their reduced extinction. Optical CL [\ion{Fe}{vii}] was also selected because of its strength, still moderate extinction, and IP close to that of [Si\,{\sc vi}].

The final working sample of objects has 31 AGN (Table~\ref{tab:data}). It includes BH masses, most  from \citet{bentz+15} compilation, coronal line ratios measured from spectra in this work  or from spectra already described in other publications, and data sources. For a sub-sample,  optical and near-IR data are presented in this work for the first time.  Table~\ref{tab:dados} lists these later sources with details of the observations, including telescope/instrument employed, date of observation, airmass and exposure time.

\subsection{Optical Spectroscopy}

Optical spectra were taken from a variety of sources, as indicated in the last column of Table~\ref{tab:data}. In more than half of the sample, spectra from the Sloan Digital Sky Survey (SDSS) data release~7 \citep{abazajian/2009} were employed. SDSS delivers fully wavelength and flux calibrated spectra. Therefore, data reduction for these objects will not be discussed here. Similarly, archival flux-calibrated spectra for Ark\,564 taken by the Faint Object Spectrograph (FOS) on-board the Hubble Space Telescope (HST) were employed. Details of observations and reduction of this target will not be addressed here. NGC\,4051 employs archival spectroscopy available from the NASA's Extragalactic Database (NED). Details of that observation can be found in \citet{moustakas+06}. Mrk~335 was observed using the 2.15~m telescope at the Complejo Astron\'omico El Leoncito (CASLEO). Details of the observations and data reduction are in \citet{rodriguez+02}. Fully reduced spectra for Fairall~9, NGC~4151, and Mrk~509 were extracted from the AGN Watch Project\footnote{\href{http://www.astronomy.ohio-state.edu/~agnwatch/}{http://www.astronomy.ohio-state.edu/~agnwatch/}}.

The second major source of optical data is the 4.1 m Southern Observatory for Astrophysical Research (SOAR) Telescope at Cerro Pachon, Chile. The observations were carried out using the Goodman Spectrograph \citep{clemens/2004}, equipped with a 600~l/mm grating and a 0.8 arcsec slit width, giving a resolution R$\sim$1500. In addition to the science frames, standard stars \citep{baldwin/1984} were observed for flux calibration. The average seeing of the observations was $\sim$ 1 arcsec. None of the nights were photometric, implying that the flux calibration is relative. Accordingly,  no attempt  was made to put the optical and NIR data on the same flux scale. Hg-Ar arc lamps were taken after the science frames for wavelength calibration. Daytime calibrations include bias and flat field images.
 
The data were reduced using standard {\sc iraf} tasks. It includes subtraction of the bias level and division of the science and flux standard star frames by a normalised master flat-field image. Thereafter, the spectra were wavelength calibrated by applying the dispersion solution obtained from the arc lamp frames. Finally, the spectra of standard stars were extracted and combined to derived the sensitivity function, later applied to the 1D science spectra. The final products are wavelength and flux calibrated optical spectra.

In all cases above, the final spectra were corrected for Galactic extinction using the extinction maps of \citet{schlafly/2011} and the \citet{cardelli/1989} extinction law.  Figures~\ref{fig:optspec1} to~\ref{fig:optspec4} show the optical spectra employed in this work.

\subsection{NIR Spectroscopy}

Most of the NIR emission line flux ratios employed in this work were extracted from \citet{riffel+06}. For targets not reported in that publication, observations were obtained using either the Gemini Near-Infrared Spectrograph (GNIRS) attached to the Gemini North Telescope or the ARCOiRIS spectrograph, mounted on either Blanco or SOAR Telescopes. Note that ARCOiRIS was installed on Blanco since 2017 and up to 2019, when it was then moved to SOAR with no modifications regarding their setup. Below we describe the observations and data reduction procedures, noting that no distinction between Blanco and SOAR is made. Both data collection and treatment is made employing the same observing strategy and reduction pipelines.

Figures~\ref{fig:nirspec1} to~\ref{fig:nirspec3} show the final reduced NIR spectra for those galaxies not reported in~\citet{riffel+06}.

\subsubsection{ARCOiRIS Blanco/SOAR data}

NIR spectra of Fairall~9, 3C~120, Mrk~707, NGC\,3783, Mrk\,1310, Mrk\,841 and NGC\,6814 were obtained using the ARCoIRIS spectrograph attached to either the 4\,m Blanco Telescope or the 4.1 m SOAR telescope. The science detector employed is a 2048 $\times$ 2048 Hawaii-2RG Hg-Cd-Te array with a sampling of 0.41 arcsec/pixel. The slit assembly is 1.1 arcsec wide and 28 arcsec long. The delivered spectral resolution R is $\sim$3500 across the different dispersion orders. Observations were done nodding in two positions along the slit. Right before or after the science target, a telluric star, close in airmass to the former, was observed to remove telluric features and to perform the flux calibration. Cu-Hg-Ar frames were also observed at the same position as the galaxies for wavelength calibration.

The spectral reduction, extraction and wavelength calibration procedures were performed using {\sc spextool v4.1}, an IDL-based software developed and provided by the SpeX team \citep{cushing/2004}  with some modifications specifically designed for the data format and characteristics of ARCoIRIS, written by Dr. Katelyn Allers (private communication). Telluric features removal and flux calibration were done using {\sc xtellcor} \citep{vacca/2003}. The different orders were merged into a single 1D spectrum from 1 to 2.4~$\mu$m using the {\sc xmergeorders} routine. We then corrected these data for Galactic extinction using the \citet{cardelli/1989} law and the extinction maps of \citet{schlafly/2011}.

\subsubsection{GNIRS/Gemini spectroscopy}

Near-infrared spectra of NGC\,4395 and Ark\,564 were collected using
GNIRS \citep[][]{elias/2006} in the cross-dispersed mode. It allows simultaneous z+J, H and K band observations, covering the spectral range 0.8\,$-$\,2.5$\mu$m in a single exposure. GNIRS science detector consist of an ALADDIN 1k $\times$ 1k In-Sb array. The instrument setup includes a 32~l/mm grating and a 0.8$\times7$ arcsec slit, giving a spectral resolution of R$\sim$1300 (or 320~km\,s$^{-1}$ FWHM). Individual exposures were taken, nodding the source in a ABBA pattern along the slit.
Right after the observation of the science frames, an A0V star was observed at a similar airmass, with the purpose of flux calibration and telluric correction.
 
The NIR data were reduced using the XDGNIRS pipeline (v2.0)\footnote{Based on the Gemini IRAF packages}, which delivers a fully reduced, wavelength and flux calibrated, 1D spectrum with all orders combined \citep{mason/2015}. Briefly, the pipeline cleans the 2D images from radiative events and prepares a master flat constructed from quartz IR lamps to remove pixel to pixel variation. Thereafter, the s-distortion solution is obtained from daytime pinholes flats and applied to the science and telluric images to rectify them. Argon lamp images are then used to find the wavelength dispersion solution, followed by the extraction of 1D spectra from the combined individual exposures. The telluric features from the science spectrum are removed using the spectrum of a A0V star. Finally, the flux calibration is achieved assuming a black body shape for the standard star \citep{pecaut/2013} scaled to its $K$-band magnitude \citep{skrutskie/2006}. The different orders are combined in to a single 1D spectrum and  corrected for Galactic extinction using the \citet{cardelli/1989} law and the extinction maps of \citet{schlafly/2011}.  
 The spectra of these two sources are not presented here as they are shown in \citet{mason/2015}.

In order to measure the flux of the lines for the sub-sample of objects described above, we modelled the observed profiles with a suitable function that best represents them and then integrated the flux under that function. To this purpose we employ the {\sc liner} routine (\citealt{pogge/1993}).  This software performs a least-square fit of a model line profile (Gaussian, Lorentzian, or Voigt functions) to a given line or set of blended lines to determine the flux, peak position and FWHM of the individual components. Typically, one or two Gaussian components were necessary to represent the coronal lines. For the permitted lines of \ion{H}{i} a broad component associated to the BLR was employed. In this process, the underlying continuum emission was approximated by a linear fit. 

For the optical part, the measurement of the H$\beta$ flux was preceded by the removal of the underlying power-law continuum and the pseudo-continuum produced by the \ion{Fe}{ii} lines that contaminates H$\beta$. This was done following the prescription of \citet{bg92}.

Table~\ref{tab:data} shows the measured optical emission line flux ratio between [\ion{Fe}{vii}]~$\lambda$6087 and the broad component of H$\beta$ (column~3) and the NIR ratios for [\ion{Si}{vi}]~1.964~$\mu$m/Br$\gamma$ (column 4), [\ion{Si}{x}]~1.431~$\mu$m/Pa$\beta$ (column 5), and [\ion{S}{viii}]~0.9914~$\mu$m/Pa$\beta$ (column 6). For the later three ratios the flux associated to the broad component of the Brackett or Paschen lines was employed. Table~\ref{tab:fluxes} in the Appendix lists the individual fluxes of all lines employed in this work. Note that because the optical and NIR spectra were taken on different dates and in most cases, different telescopes, the intrinsic line ratios H$\beta$/Pa$\beta$ and H$\beta$/Br$\gamma$ may depart from their theoretical value. This, however, does not affect our results as we do not use   line ratios that combine both spectral regions. However, for consistency, the reported  fluxes for the broad component of H$\beta$, Pa$\beta$ and Br$\gamma$ are compared to those reported by \citet{landt08} for the objects in common. We found the values agree within a factor of 2, with most cases the difference being  not larger than 30\%. Moreover, our measured fluxes for NGC\,4151 and NGC\,5548 are in excellent agreement to those reported by \citet{landt15a} and \citet{landt15b} for these two objects, respectively.

\begin{table*}
\centering
\caption{Black hole mass and Coronal Line ratios for the galaxy sample.}
\label{tab:data}
\begin{tabular}{lccccc}
\hline\hline\noalign{\vskip 0.1cm}
\multicolumn{1}{l}{Galaxy} & log $M_{\rm BH}^1$ & [\ion{Fe}{vii}]/H$\beta ^{2}$ & [\ion{Si}{vi}]/Br$\gamma^{3}$ & [\ion{Si}{x}]/Pa$\beta$ & [\ion{S}{viii}]/Pa$\beta$ \\

\hline\noalign{\vskip 0.1cm}
						
Mrk\,335 & 7.23$\pm$0.04 & 0.077$\pm$0.009 & 0.40$\pm0.09^{e}$ & 0.04$\pm0.01$ & 0.022$\pm0.006$  \\
Fairall\,9 & 8.29$\pm$0.09 & 0.038$\pm$0.003 & 0.11$\pm0.02^{f}$ & 0.08$\pm0.01$ & 0.04$\pm0.01$  \\
NGC\,863 & 7.57$\pm$0.06 & ... & 0.24$\pm0.07^{e}$ & ... & ...\\
3C\,120 & 7.74$\pm$0.04 & ... & 0.33$\pm0.06^{f}$ & 0.08$\pm$0.02 & 0.03$\pm$0.01 \\
Mrk\,707 & 6.50$\pm$0.10$^a$ & 0.025$\pm$0.002 & 0.38$\pm0.04^{f}$ & ... & ... \\
Mrk\,110 & 7.29$\pm$0.10 & 0.05$\pm$0.002 & ... & ... & ... \\
NGC\,3227 & 6.78$\pm$0.10 & ... & 0.75$\pm$0.20 & ... & 0.012$\pm$0.004 \\
Mrk\,142 & 6.29$\pm$0.10 & 0.02$\pm$0.004 & ... & ... & ... \\
SBS\,1116+583A & 6.56$\pm$0.09 & 0.01$\pm$0.002 & ... & ... & ...  \\
PG\,1126-041 & 8.08$\pm$0.03$^b$ & ... & 0.28$\pm$0.02 & 0.04$\pm$0.01 & 0.025$\pm$0.002  \\
NGC\,3783 & 7.37$\pm$0.08 & 0.052$\pm$0.002 & 0.42$\pm0.09^{f}$ & 0.05$\pm$0.01 & 0.023$\pm$0.003  \\
Mrk\,1310 & 6.21$\pm$0.08 & 0.032$\pm$0.002 & 0.57$\pm0.17^{f}$ & 0.06$\pm$0.01 & 0.06$\pm$0.01 \\
NGC\,4051 & 6.13$\pm$0.12 & 0.123$\pm$0.012 & 0.96$\pm$0.11 & 0.33$\pm$0.02 & 0.205$\pm$0.030  \\
NGC\,4151 & 7.55$\pm$0.05 & 0.02$\pm$0.001 & 0.51$\pm$0.05 & 0.05$\pm$0.01 & 0.057$\pm$0.003  \\
Mrk\,202 & 6.13$\pm$0.17 & 0.033$\pm$0.004 & ... & ... & ... \\
Mrk\,766 & 6.82$\pm$0.05 & 0.03$\pm$0.002$^d$ & 0.78$\pm$0.10 & 0.05$\pm$0.01 & 0.045$\pm$0.002 \\
Mrk\,50 & 7.42$\pm$0.06 & 0.005$\pm$0.001 & ... & ... & ... \\
NGC\,4395 & 5.45$\pm$0.13 & 0.092$\pm$0.005 & 1.18$\pm0.10^{f}$ & 0.02$\pm$0.01 & 0.053$\pm$0.006  \\
Mrk\,771 & 7.76$\pm$0.20 & 0.03$\pm$0.002 & ... & ... & ...  \\
NGC\,4748 & 6.41$\pm$0.11 & ... & 0.93$\pm$0.06 & 0.06$\pm$0.02 & 0.137$\pm$0.042 \\
PG\,1307+085 & 8.54$\pm$0.13 & 0.01$\pm$0.002 & ... & ... & ... \\
MGC-6-30-15 & 6.60$\pm$0.12 & 0.017$\pm$0.002 & ... & ... & ...\\
NGC\,5548 & 7.72$\pm$0.02 & 0.044$\pm$0.003 & 0.61$\pm$0.09 & 0.11$\pm$0.01 & 0.122$\pm$0.011 \\
PG1448+273 & 6.97$\pm$0.08 & ... & 0.57$\pm$0.11 & ... & ... \\
Mrk\,290 & 7.28$\pm$0.02 & 0.023$\pm$0.002 & ... & ... & ... \\
Mrk\,841 & 8.10$\pm$0.02$^c$ & 0.008$\pm$0.002 & 0.20$\pm0.06^{f}$ & 0.02$\pm$0.01 & 0.024$\pm$0.002 \\
3C\,390.3 & 8.64$\pm$0.04 & 0.018$\pm$0.001 & ... & ... & ... \\
NGC\,6814 & 7.04$\pm$0.06 & ... & 0.13$\pm0.02^{f}$ & ... & ...\\
Mrk\,509 & 8.05$\pm$0.04 & ... & 0.17$\pm$0.02 & ... & ...  \\
Ark\,564 & 6.59$\pm$0.17 & 0.056$\pm$0.007 & 1.08$\pm$0.10 & 0.30$\pm$0.01 & 0.101$\pm$0.006 \\
NGC\,7469 & 6.96$\pm$0.05 & 0.021$\pm$0.001 & 0.60$\pm$0.05 & 0.07$\pm$0.01 & 0.037$\pm$0.006 \\
\hline\\ 
\end{tabular}

{
\begin{minipage}{0.9\textwidth}
All the line ratios are normalised to the broad component of HI. Individual line fluxes are  listed in Table~\ref{tab:fluxes}
 1. Masses are from \citet{bentz+15} unless stated otherwise; (\textit{a}) -- \citet{park+17}; (\textit{b}) -- \citet{dasyra+07}; (\textit{c}) -- \citet{woo+02}. 2. [\ion{Fe}{vii}] ratio is   determined in this work unless stated otherwise: (\textit{d}) -- \citet{rodriguez+05}. 3. The NIR emission line flux ratios are from \citet{riffel+06} except when indicated: (\textit{e}) -- \citet{rodriguez+02}; (\textit{f}) -- This work.     
\end{minipage}    
}
\end{table*}

\begin{table*}
\caption{Sub-sample of galaxies with observations first presented in this work}
\resizebox{\textwidth}{!}{%
\begin{tabular}{lccccccccc}
\hline
\multicolumn{1}{c}{Galaxy} & RA & DEC & Redshift & Telescope/ & Date  of     & A$_{\lambda}$V & Airmass & Nexp $\times T_{\rm exp}$ \\    
             & hh.mm.ss    & deg.mm.ss      & (z) & Instrument     & Observation   & mag  &     & (sec)    \\
\hline
Fairall\,9 & 01:23:45.8 & -58:48:21 & 0.04614 & SOAR/TSpec4 & 2019 Aug 08 & 0.071  & 1.22 & 20x180 \\
3C\,120 & 04:33:11.1 & 05:21:16 & 0.03301 & SOAR/TSpec4 & 2020 Feb 10 & 0.816   &  1.28 & 12x180 \\
Mrk\,707 & 09:37:01.03 & 01:05:43.48 & 0.05025 & Blanco/ARCOiRIS & 2017 Apr 09 & 0.189  & 1.21 & 28x180 \\
         &             &             &         & SOAR/Goodman & 2017 Mar 11 &  & 1.18 & 3x900 \\
NGC\,3783 & 11:39:01.7 & -37:44:19 & 0.00973 & Blanco/ARCOiRIS & 2017 Apr 10 & 0.332  & 1.05 & 16x180 \\
         &             &             &         & SOAR/Goodman & 2011 Jan 31 &  & 1.10 & 3x900 \\
Mrk\,1310 & 12:01:14.3 &  -03:40:41 & 0.01956 & Blanco/ARCOiRIS & 2017 Apr 08 & 0.083  & 1.18 & 20x180 \\
         &             &             &         & SOAR/Goodman & 2012 May 21 &  & 1.24 & 3x1800 \\
MGC-6-30-15 & 13:35:53.7 & -34:17:44 & 0.00749 & SOAR/Goodman & 2011 Apr 01 & 0.165 & 1.05 &  3x1200  \\
Mrk\,841 & 15:04:01.2 & 10:26:16.15 & 0.03642  & Blanco/ARCOiRIS & 2017 Apr 08 & 0.082  & 1.42 & 12x180 \\
         &             &             &         & SOAR/Goodman & 2017 Mar 11 & & 1.36   & 2x900 \\
NGC\,6814 & 19:42:40.6 & -10:19:25 & 0.00521 & Blanco/ARCOiRIS & 2017 Apr 10 & 0.509 & 1.38 & 20x180 \\
NGC\,7469 & 23:03:15.6 & +08:52:26 & 0.01632 & SOAR/Goodman & 2010 Nov 11 & 0.188 & 1.34 & 3x900 \\
\hline
\end{tabular}}
\label{tab:dados}%
\end{table*}

\section{Coronal line diagnostic diagrams}
\label{sec3}
 
The use of coronal lines as a proxy of black hole measurements was theoretically explored by \citet{cann+18} via photoionisation simulations. They show that for intermediate to low BH mass sources ($10^2 - 10^5$~M$_\odot$), CL of very high IP are favoured with respect to those of lower IP. Their approach, though, could not be assessed because of the lack of suitable data for AGNs with BH masses in that regime. Here, we  expand the photoionisation modelling to the high BH mass range $10^6 - 10^8$~M$_\odot$, and confront the predictions with the data gathered in Table 1.

\begin{figure*}
	\includegraphics[scale=0.8]{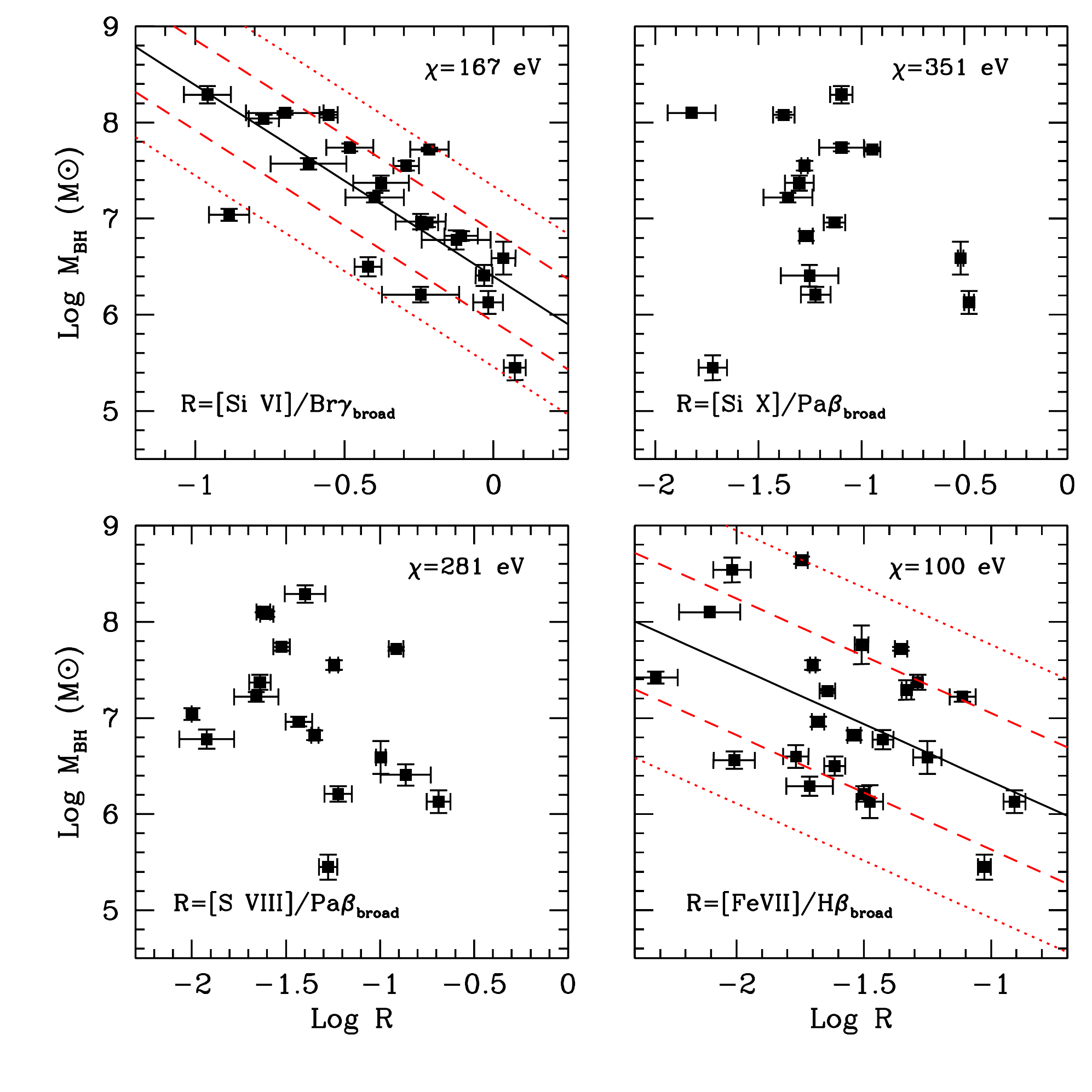}
    \caption{Observed CL emission normalised to the broad component of H\,{\sc ii} versus black hole mass for the objects in sample.  The black  line is the best linear fit to the data and the red-dashed and -dotted lines show the 1$-\sigma$ and 2$-\sigma$ deviation.
    } 
    \label{fig:ratios}
\end{figure*}

Fig.~\ref{fig:ratios} presents new diagnostic  diagrams in which the BH mass for the objects  in the sample is plotted against a given CL flux normalised to the closest in wavelength H\,{\sc i} broad emission. The first plot - upper left panel, involving [\ion{Si}{vi}]~1.9641~$\mu$m/Br$\gamma_{\rm broad}$, IP [\ion{Si}{vi}] = 167 eV,  shows a clear trend with $M_{\rm BH}$ over three orders of magnitude in BH mass. A linear regression yields:
\begin{equation}
    \log M_{\rm BH} = (6.40\pm 0.17) - (1.99\pm 0.37) \times \log \left(\rm{\frac{[Si\,VI]}{Br\gamma_{\rm broad}}}\right),
\end{equation}
and a 1$\sigma$ dispersion of 0.47 dex in BH mass. The regression analysis follows the {\sc LtsFit} package\footnote{\href{http://www-astro.physics.ox.ac.uk/~mxc/software/\#lts}{http://www-astro.physics.ox.ac.uk/mxc/software/lts}} \citep{capellari+13}, which accounts for the errors in all variables. The Pearson correlation coefficient is $r$ = -0.76, with a null probability of Pr = 3.8$\times 10^{-5}$. 

A weak trend if any (\textit{r} = -0.5) when involving [\ion{Fe}{vii}], IP (100 eV), and no trend for the higher IP ($>$ 260~eV) CL [\ion{Si}{x}] and [\ion{S}{viii}] are found. The correlation index for [\ion{S}{viii}]/Pa$\beta_{\rm broad}$ is $r$ = -0.44, that for [\ion{Si}{x}]/Pa$\beta_{\rm broad}$, is $r$= -0.3.

To construct these diagrams, several considerations were made. First, line ratios are chosen close in wavelength to minimise reddening. Second, the normalisation of the CL emission is to the broad \ion{H}{i} line, as only in this case the correlation  ([\ion{Si}{vi}]/Br$\gamma_{\rm broad})$ with BH mass was found. Normalisation to the \ion{H}{i} narrow line emission was equally evaluated, yet not  clear trend with BH mass could be recovered, the dispersion being too large. We believe that reasons for that  are the much larger  narrow-\ion{H}{i} emitting volume as compared with that  of the coronal region 
and the fact that H is subjected to  additional ionisation sources beside  the AGN, e.g.  star formation, whereas  coronal gas is an unambiguous AGN tracer. The coronal gas is found in the inner parsecs of the central engine, at the boundary of the  BLR clouds, as recently shown in spatially resolved  observations of the broad and coronal clouds by \cite{gravity2021}. The CL high critical density, $n_{\rm e}> 10^8$ cm$^{-3}$, warrants its survival at the inner regions, possibly the nearest gas to   the BLR - not at the BLR where densities are  at least one order of magnitude higher. Hence, by  normalising to broad \ion{H}{i}, the proposed CL ratios  in this work become  the closest possible tracer, perhaps a genuine one,  of the conditions at the inner parsecs next to  the accretion disc. 

The derived correlation  when involving the $[\ion{Si}{vi}]$ line over almost three order of magnitude in BH mass, the absence of an equivalent dependence when involving the higher IP lines, prompted us to examine the impact of an accretion-disc continuum in the coronal  gas production and, in turn, its dependence with disc temperature and BH mass.

\begin{figure}
     \includegraphics[width=\columnwidth]{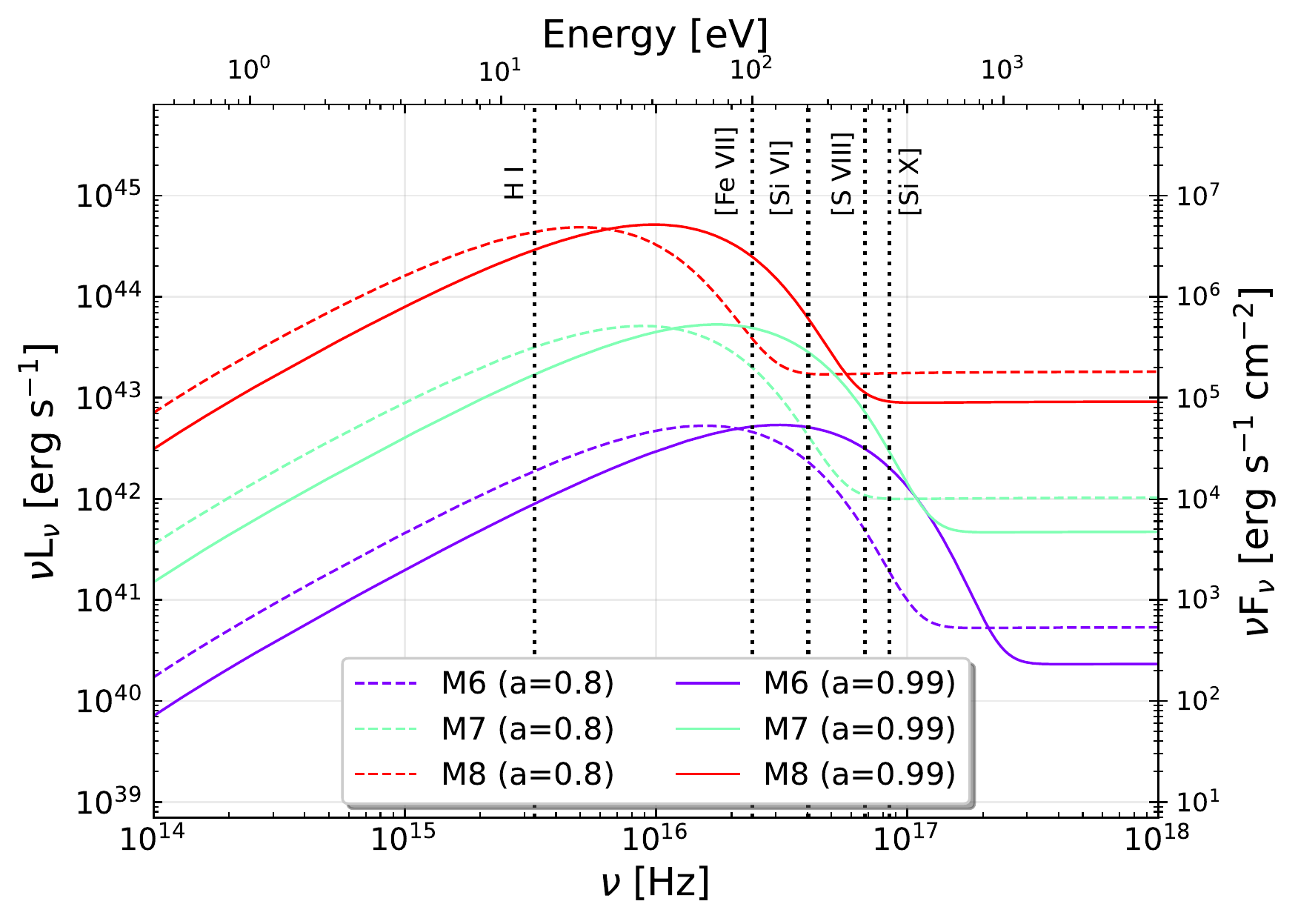}
     \caption{Generic AGN ionising continuum used in this work as per Eq. \ref{eq:sed} for the BH mass range in the AGN sample. Each curve corresponds to a  BH mass with an associated Tdisc following the SS disc approximation  in Eq. \ref{eq:tbbb}.Two spin values, a = 0.8 (dashed curves) and a = 0.99 (solid curves), and three BH masses, i.e. \mbh{} (in units of \msun{}) = 10$^6$ (purple), 10$^7$ (green) and 10$^8$ (red) are shown. Vertical dashed lines mark the IPs of the lines used in the analysis.}
   \label{fig:continuum}
\end{figure}

 The CL in Fig. 1 are sensitive to different energy ranges of the ionising continuum. This is illustrated in Fig~\ref{fig:continuum}, which shows a parameterisation of an AGN ionising continuum as a combination of a standard Shakura-Sunyaev (SS) accretion disc \citep{ss73} which accounts for the rising of the spectrum at UV-soft X-rays, and a power-law with a low and high energy cut-off to account for the rising of the continuum at high energies. It follows the equation (CLOUDY C13.1 formalism is used  - Ferland et al. 2013): 

\begin{equation}
    F_{\nu} = \nu^{\alpha_{uv}}\exp{\left(\frac{-h\nu}{kT_{disc}}\right)}\exp{\left(\frac{-kT_{IR}}{h\nu}\right)} + a\nu^{\alpha_x}
\label{eq:sed}    
\end{equation}

The first term is the parameterisation of the SS disc, represented by an exponential function with a cutoff at the disc effective temperature, $T_{\rm{disc}}$, and a power law with $\alpha_{\rm uv}$ = 0.33 accounting for the low energy tail of the disc.  The low energy limit of the disc is set by the IR-exponential with cutoff at 0.01 Ryd. The high energy range is represented by a broken power law with spectral index $\alpha_{\rm x}$= -1, and a cutoff at 100 keV. The scaling of the SS disc relative to the high energy   power law is controlled by the parameter "a" in eq. 2, which refers to the ratio of the luminosities at 2 keV and at 2500 A given by   a power law with  spectral index, so called  $\alpha_{ox} $,  set here to -1.4. With the exception of $\alpha_{\rm uv}$ = 0.33, all other parameters quoted above follow the generic AGN continuum used in    CLOUDY C13.1- sect. 6.2.

The IP of the CL and of H\,{\sc ii}  are marked on the ionising continuum, Fig. 2,   to illustrate their both location and  energy coverage in a  typical AGN  ionising  spectrum.  They sample the bulk of the ionising spectrum over the 13.6 - 351 eV energy range. The figure shows different SS accretion disc continuum for different  BH masses  and  spin (see Sec. \ref{accretion-disc}). 

\section{Coronal  emission as proxy of the black hole mass}
\label{accretion-disc}

The peak emission in the ionising continuum in Fig. \ref{fig:continuum} 
effectively samples  $T_{\rm{disc}}$, with the disc getting hotter as BH mass decreases as predicted by standard accretion disc theory.  The IPs of the CL fall within the range of peak temperatures, closer or slightly further from the peak depending on BH mass. If the accretion disc is the main source of photons to ionise the coronal gas, a trend between CL strength and $T_{\rm{disc}}$ may be expected. Because of the dependence of $T_{\rm disc}$ with BH mass, a correlation between coronal line strength and BH mass may be ensued. This  possibility is investigated below.

Following on the thin accretion disc approximation, for a Kerr BH, the disc $T_{\rm disc}$ can be approximated  as (\cite{frank+02} formalism is used):

\begin{equation}
\begin{split}
 T_{\rm disc} = & 3.4
 \times 10^5 K ~ \left(\frac{M_{\rm BH}}{10^8 M_{\odot}}\right)^{1/4} \times \left(\frac{\left(\frac{dM}{dt}\right)}{0.1}\right)^{1/4}\\
 & \times \left(\frac{\eta}{0.26}\right)^{-1/4} \times \left(\frac{Rin_{G}}{1.4}\right)^{-3/4}
 \label{eq:tbbb}
\end{split}
\end{equation}

where, $M_{\rm BH}$ is the BH mass, $(dM/dt)$ is the accretion rate in Eddington units, $\eta$ is the BH accretion efficiency, $Rin_{G}$ is the inner-most stable circular orbit  in terms of the gravitational radius $R_G = GM_{\rm BH}/c^2$ (\textit{G} is the Gravitational constant, \textit{c} is the velocity of light).
The equation is normalised to  $M_{\rm BH}$ = 10$^8$ M$_{\odot}$, accretion rate in Eddington units,  $dM/dt$ $\sim$ 0.1 $(dM/dt)_{Edd}$, and a radiation efficiency, $\eta$ = 0.26 corresponding to a BH spin of 99\% , i.e.   $ a = 0.99 ~GM_{\rm BH}/c^2$ (hereafter  a = 0.99), and co-rotation is adopted. Current estimates of BH spins from different methods  point to values close to 1 \citep[see][for a compilation]{reynolds2019}. With increasing spin, the inner-most stable orbit becomes smaller and $T_{\rm disc}$ increases accordingly. In the case of co-rotation, the disc temperature reaches the highest values.

The ionising continua in Fig.~\ref{fig:continuum} follow the $T_{\rm disc}$ - BH mass approximation described in Equation \ref{eq:tbbb}. Curves are presented for  three different BH mass  and two spin, 0.8 and 0.99.  Spins below 0.8 produce a marginal difference in $T_{\rm disc}$ in comparison to the $a$ = 0.8 case and are  not shown \citep[see e.g.][for a spin parameter space study]{campitiello+19}.
It can be seen that the peak of the disc emission gets progressively closer to the IP  of $Si^{+5}$ and $Fe^{+6}$  as we move down in BH mass, and  increase  spin. However,  the accretion disc energies begin to fall short from those required for the higher IP ions, $Si^{+9}$ and $S^{+7}$, when moving to  the high BH masses,  $10^7  - 10^8$ \msun{}, or higher - not shown - with   $T_{disc}$ becoming progressive cooler with increasing BH mass.   
 
On the above premises, a trend between $T_{\rm disc}$ and the excitation state of [\ion{Si}{vi}] or [\ion{Fe}{vii}] lines may be expected, whereas no trend involving the higher IP lines [\ion{Si}{x}] or [\ion{S}{viii}] is foreseen. The high IP lines  are however  expected to be good proxies of the disc temperature for BH mass below 10$^6$ M$_{\odot}$, as  shown in the theoretical analysis by \citet{cann+18}. We believe that the  expected  CL emission dependence  with disc temperature is a possible driver  of the observed [\ion{Si}{vi}] ~/~Br$\gamma_{\rm broad}$ -- $M_{\rm BH}$ correlation,  and  may also explain the absence of an equivalent  dependence for the higher IP [\ion{Si}{x}] or [\ion{S}{viii}] lines. The hypothesis  is tested in the next subsection with photoionisation models.

We note also that contrarily to expectations, the trend with BH mass  for  the [\ion{Fe}{vii}]  case is  weak, if any, even though the  IP of Fe$^{6+}$ is  well sampled by the ionising continua (Fig.~\ref{fig:continuum}).   We are not yet  understanding this result, but identify several observational issues that may affect the estimate of this CL ratio: 1) the   high  variability of H$\beta$ as compared with the much more stable near-IR Paschen and Brackett lines \citep{landt+11}; 2)  the continuum underlying  broad H$\beta$, affected by a strong \ion{Fe}{ii} pseudo-continuum,  makes difficult the estimate of  broad H$\beta$;  3) differential reddening between  H$\beta$ and [\ion{Fe}{vii}]. Nonetheless,  this CL ratio is included in the photoionisation modelling below.

\subsection{Testing Coronal Line diagnosis diagrams with photoionisation models} 
\label{modelling}
 
To probe the  CL emission as a proxy of the disc temperature, we make use of the photoionisation code CLOUDY \citep[v17.02][]{ferland+17}. The goal 
is to test whether the CL line ratios used in Fig. \ref{fig:ratios} show a dependence with $T_{\rm{disc}}$, this being an input to CLOUDY via the ionising continua shown in Fig. \ref{fig:continuum}. A range of electron densities, $n_{e}$, and cloud distances to the centre, \textit{r}, in line with CL observations are  input to CLOUDY. These parameters are selected as follows.

 Coronal gas  extends  at most up to a few tens of parsec \citep{prieto+05}. The bulk of the emission is mostly nuclear \citep{muller-sanchez+11}, spreading over sub-parsec scales  \citep[e.g.][]{gravity2020,gravity2021} and can be explained  by photoionisation \citep[e.g.][]{ferguson+97,contini_viegas2001,rodriguez-ardila06}. 
For the present test,  the  bulk of CL in the sample is assumed to be nuclear and powered by photoionisation from the accretion disc.  Spatially resolved nuclear CL by GRAVITY interferometry reveals the coronal region extending up to 0.3~pc from the centre. Thus,  a range of $r$ between 0.3 and 30~pc are tested.

Gas densities, $n_{\rm e}$, in the  $10^4 \leq n_{\rm e} \leq 10^7$ cm$^{-3}$  range are probed, the upper limit set by the critical density of the CL probed - $n_{\rm e} >10^8$ cm$^{-3}$, the lower limit is set by the average densities inferred from the mid-IR [\ion{Ne}{v}] lines, IP=97 eV, $10^3 - 10^4$~cm$^{-3}$  \citep[e.g.][]{moorwood_etal1996,Fernandez-Ontiverosetal2016}.

The input ionising continuum is set by the Eq. \ref{eq:sed} (Fig.~\ref{fig:continuum}), normalised to an adopted  value of \LLEdd{} = 0.1. The present sample covers a  range of Eddington ratio  in the range 0.01 $\lesssim$ \LLEdd{} $\lesssim$ 0.2 as reported in the literature. These values are nonetheless subject to uncertainties up to an order of magnitude due to the evaluation methods and assumptions made to estimate $L_{\rm bol}$ \citep[see e.g.][]{richards_etal2006}. For a few sources, reliable $L_{\rm bol}$ are available  from integrating the parsecs-scale SED  \citep[e.g.][]{prieto_etal2010}. This is the case for NGC 7469 and NGC 3783, for which \LLEdd{} are  0.25 and 0.06, respectively. A few other sources  have $L_{\rm bol}$ estimated from hard X-ray data  \citep{winter_etal2012}, re-normalising the Eddington ratios inferred by these authors to the BH masses in Table \ref{tab:data}, yields \LLEdd{} in the 0.01 - 0.1 range.  For the present test,  a nominal   \LLEdd{} = 0.1 is adopted, in line with the median  found in larger quasar distributions, such as the SDSS DR7 quasar catalogue \citep{Shen+11,panda18b}. 
 The impact  of changing \LLEdd{} in the  model results is discussed  in next section.
 
\begin{figure*}
	\includegraphics[scale=1.1]{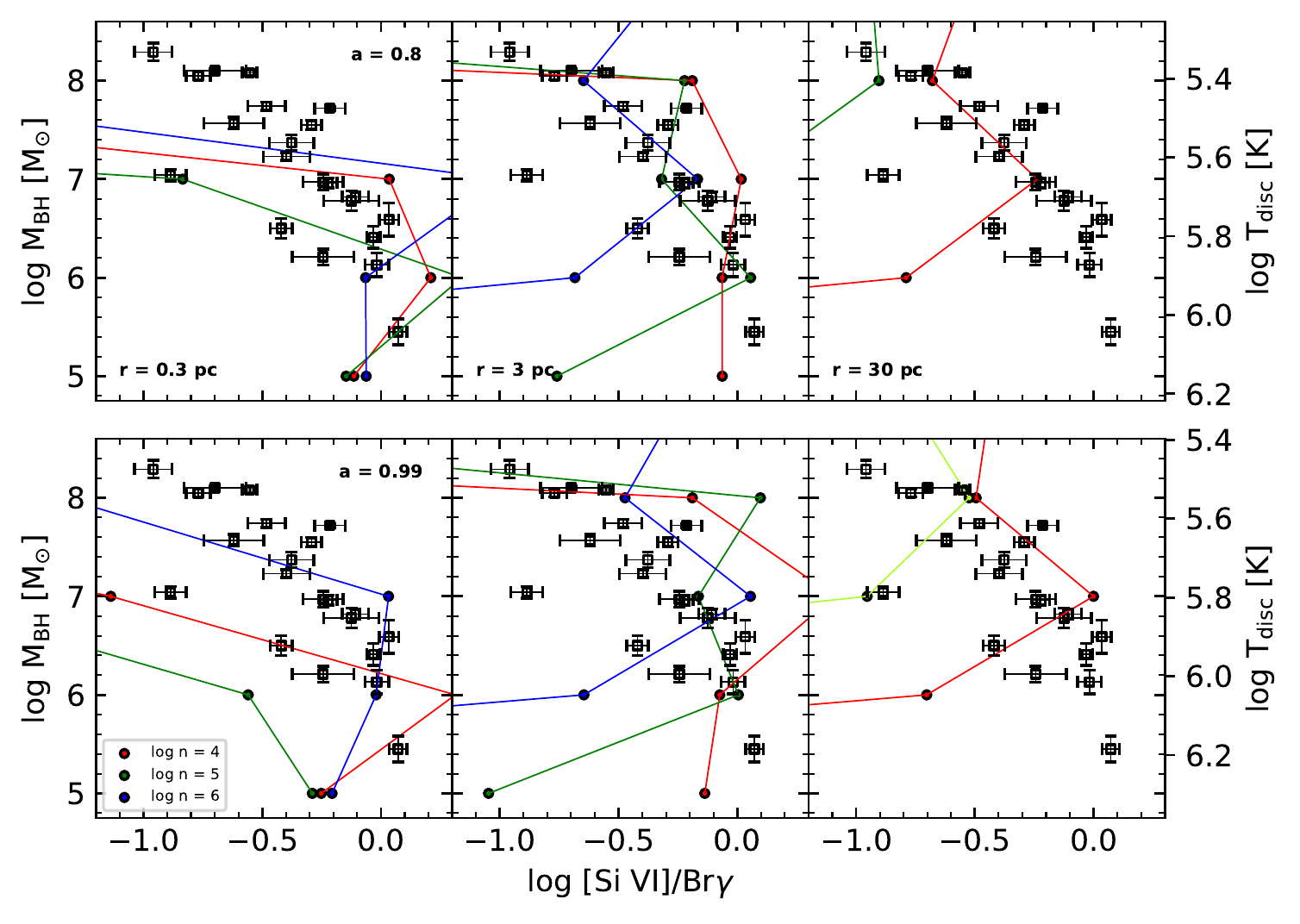}
    \caption{CLOUDY predictions for [\ion{Si}{vi}]/Br$\gamma_{\rm broad}$ vs BH mass  using as ionising continuum  Eq. \ref{eq:sed}. Models are run for densities ne = $10^3 cm^{-3}$ in red, $10^5 cm-3$ (green) and  $10^6 cm^{-3}$ (blue).
    Each subplot shows the model results for distance r to the ionising source, $r$=0.3, 3 and 30 pc,   depicted from left to right. Two spins,  a = 0.8 (upper panel) and a = 0.99 (lower panel) are considered. The corresponding $T_{\rm disc}$ per each BH mass - after eq. 3 -  are depicted on the  y-axis, right-side. Data points as in Fig. \ref{fig:ratios} are black squares.}
    \label{fig:si6brg}
\end{figure*}

\begin{figure*}
	\includegraphics[scale=1.1]{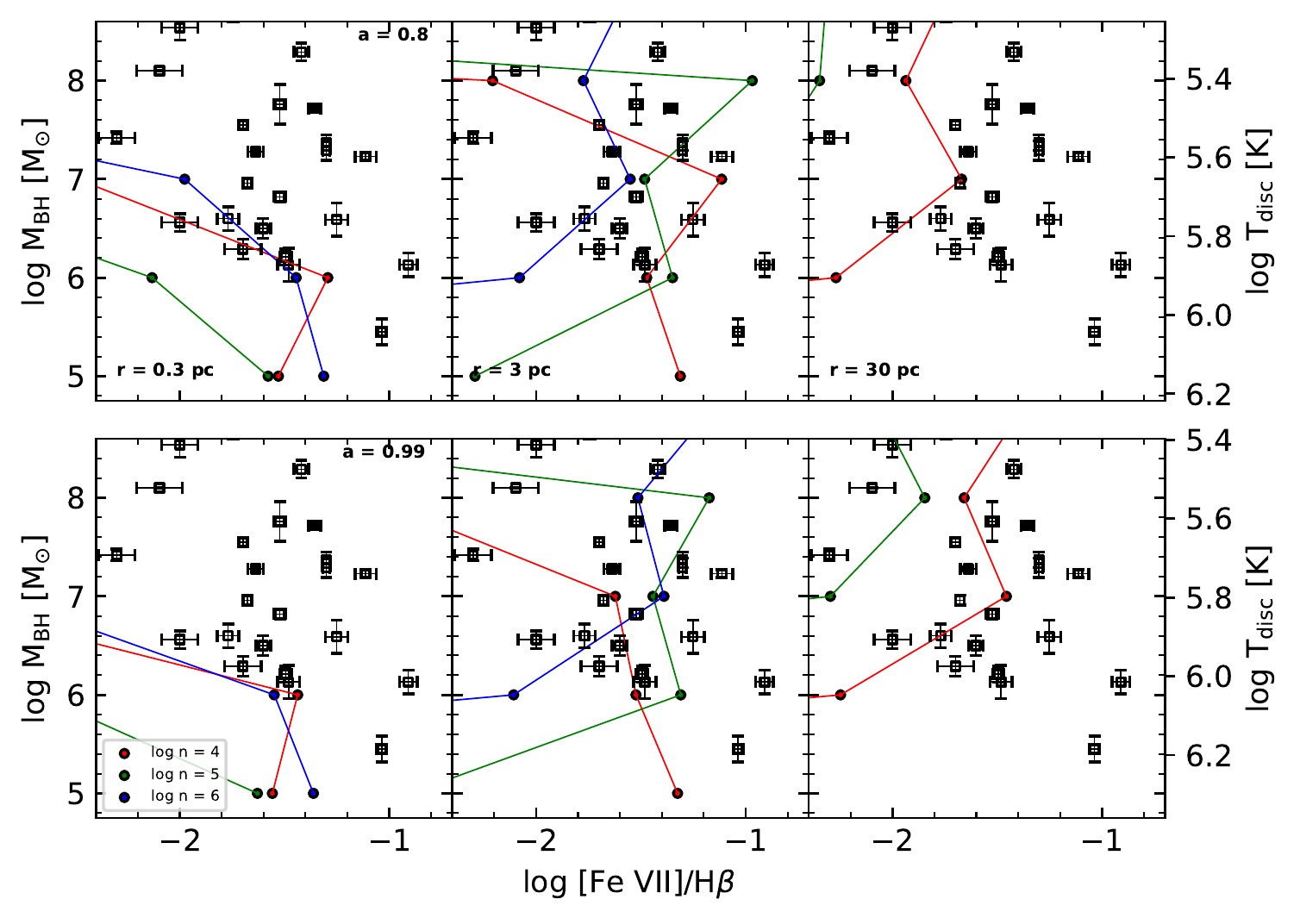}
    \caption{As in Fig. \ref{fig:si6brg} but for  [\ion{Fe}{vii}]/H$\beta_{\rm broad}$.}
    \label{fig:fe7hb}
\end{figure*}

\begin{figure*}
	\includegraphics[scale=1.1]{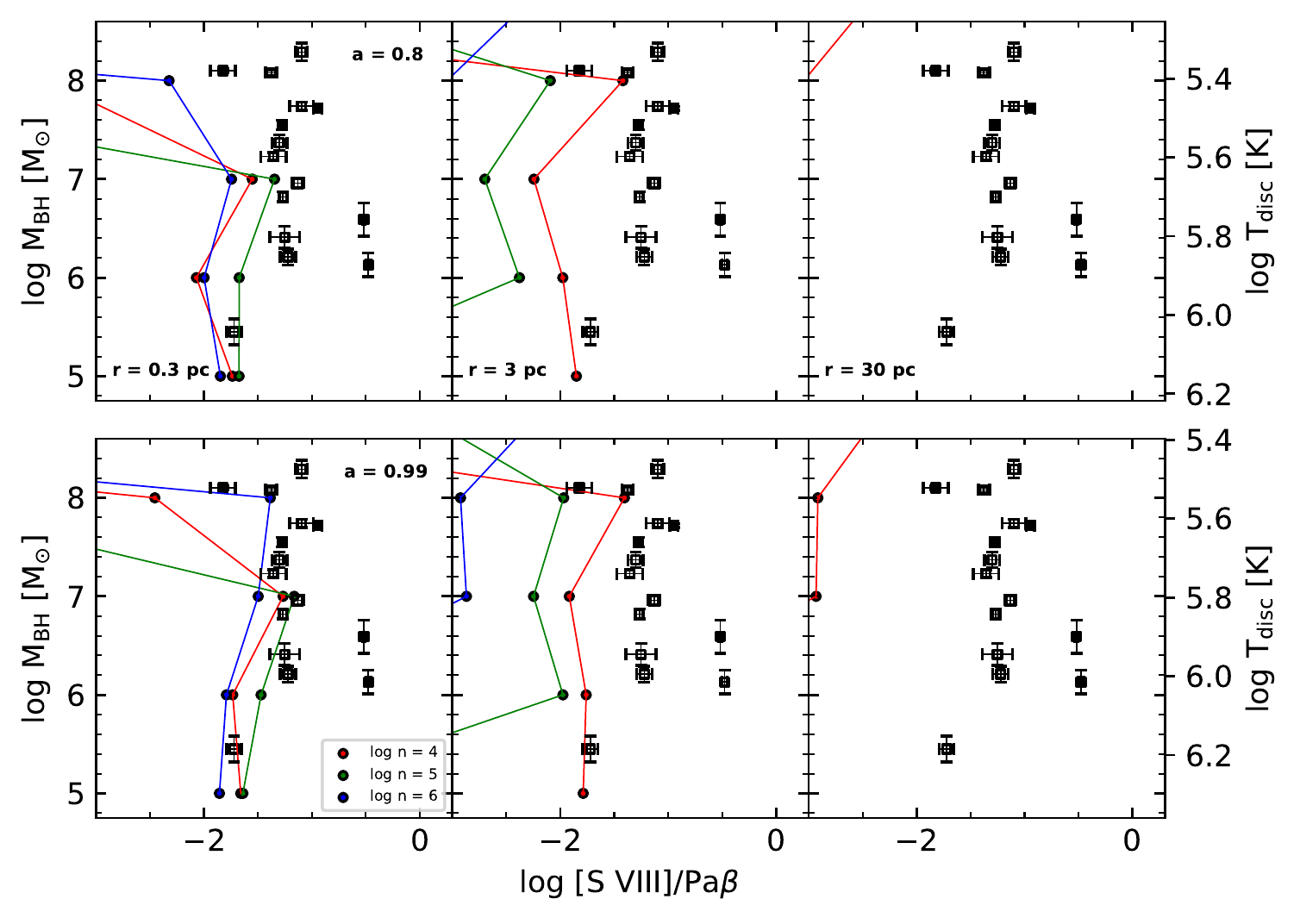}
    \caption{Similar to Fig. \ref{fig:si6brg} but for   [\ion{S}{viii}]/Pa$\beta_{\rm broad}$.}
    \label{fig:s8pab}
\end{figure*}

\begin{figure*}
	\includegraphics[scale=1.1]{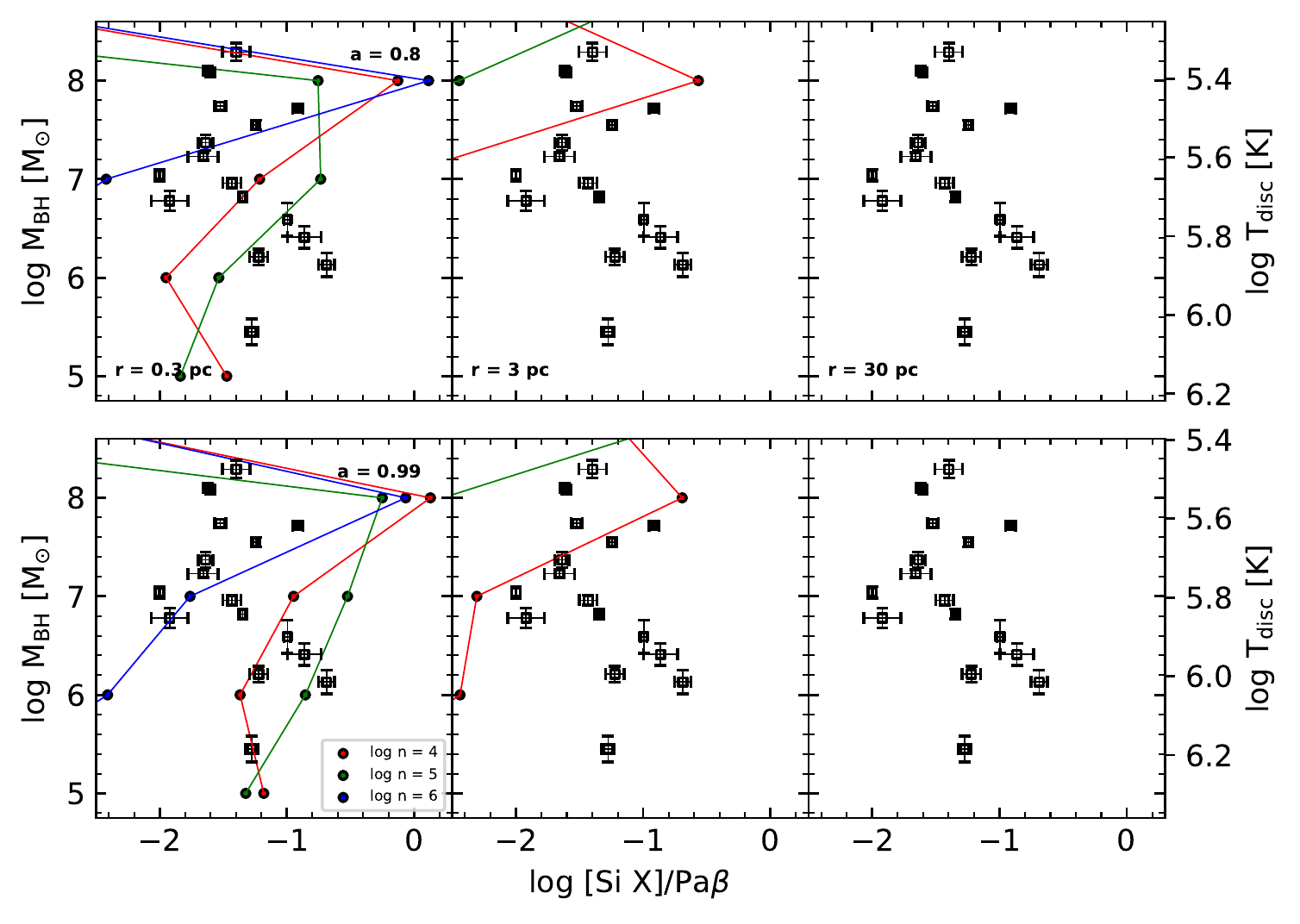}
    \caption{Similar to Fig. \ref{fig:si6brg} but for  [\ion{Si}{x}]/Pa$\beta_{\rm broad}$.}
    \label{fig:si10pab}
\end{figure*}

\subsection{Predictions vs observations}

Figs.~\ref{fig:si6brg} to~\ref{fig:si10pab} show CLOUDY predictions for the CL ratios in
this work as a function of $T_{\rm disc}$ (in the right axis), and
of BH mass  (on the left axis, following the transformation in Eq. 3 for \LLEdd{} = 0.1).
Models are shown   for a range of densities, $10^{4} \leq n_{\rm e} \leq 10^{6}$, clouds distance to the centre, 0.3 pc $\le  r \leq$ 30 pc, and  two spins, 0.8 and 0.99 - following discussion in Sect. 4.1.
Given the BH spin,  the two left free parameters in Eq. 3, accretion efficiency ($\eta$) and the location of the innermost stable circular orbit $Rin_{G}$,  are  set uniquely.
 
As discussed, the CL data is normalised to H\,{\sc i} broad emission. Yet, CLOUDY predictions are derived for densities at least two orders of magnitude below that of the broad line region to cope with the lower CL critical densities. Thus, in comparing both  a caveat is introduced, which  mainly relates to the different volume emissivity of \ion{H}{i} in the broad and in the coronal line region. To account for the difference  a correction factor was required  to shift CLOUDY line ratio predictions on top the plotted ratios. We find a factor of 15 as best compromise to  get the predicted [\ion{Si}{vi}]/Br$\gamma$, for the whole range of n$_{\rm e}$, r and spin considered, on top of the plotted [\ion{Si}{vi}]/Br$\gamma_{\rm broad}$ data. Having fixed the scale factor for this ratio,  the scaling for the other CL ratios were derived by imposing theoretical \ion{H}{i} recombination ratios. Accordingly, a factor of 25 is applied to CLOUDY's [\ion{Fe}{vii}]/H$\beta$, and 90 for both [\ion{S}{viii}]/Pa$\beta$ and [\ion{Si}{x}]/Pa$\beta$. CLOUDY models in Figs. \ref{fig:si6brg}, \ref{fig:fe7hb}, \ref{fig:s8pab} and \ref{fig:si10pab} are all shifted in the X-axis by these factors.

Focusing on  the  results for [\ion{Si}{vi}]/Br$\gamma_{\rm broad}$, Fig~\ref{fig:si6brg},  the models that best account  for   the whole range of observed ratios  and  their trend with BH mass are   for   cloud distances, $r$,  of 3~pc  and densities  in the range $10^5 - 10^6 cm^{-3}$ (middle panels). The case of spin 0.8 gives the best match,   the spin 0.99 case provides an envelope of the scatter in the data. 
Models with cloud distances higher or lower than $r =$~3 pc are more restrictive in covering the whole range of   CL ratios, or BH mass. Models with distances as large as $r =$~30~pc  provide a fair account of the trend  for the larger BH masses provided the  densities are   in the lower range  as expected for clouds at those large distances. This model nonetheless is presented as a limit case as in the present context,  the CL to  BL emission comparison may not be applicable. 
Models for $r=$~0.3~pc provide moderate account for the lowest mass range.
Overall, no single model reproduces  the whole range of observed CL ratios,  but  changes in cloud density appear to reproduce the general [\ion{Si}{vi}]/Br$\gamma_{\rm broad}$ vs BH mass trend particularly  for $r =$~3 pc.
The observed correlation shows an  scatter of 0.47 dex (Fig.~\ref{fig:si6brg}, Sect. 3), and part of it should be intrinsic, reflecting the different physical properties of the objects in the sample in terms of spin, and accretion rate - evaluated bellow - and also the CL region both density and distance to the centre. 

Focusing  on [\ion{Fe}{vii}]/H$\beta_{\rm{broad}}$, the observations show a rather scatter dependence with BH mass (Fig.~\ref{fig:fe7hb}).
 It is noteworthy though   that the models that best account for  the data spread are those   that best account for the  [\ion{Si}{vi}] ratio trend, the  $r =$~3~pc models.  All other models provide a poor representation of the loci of the data. The fact that   a favoured model, $r =$~3~pc, is hinted for both  [\ion{Fe}{vii}] and [\ion{Si}{vi}] line ratios   is  somewhat expected given that    the  IP of both lines sample properly  $T_{\rm disc}$ for  the range of BH mass considered (Fig.~\ref{fig:continuum}). It is  also interesting  that  for  the   [\ion{Fe}{vii}] case, the models   show  an also  erratic dependance with BH mass as do the data, that contrasts with the better behaved trend shown for the [\ion{Si}{vi}] line case. This result and particularly the observational one  need to be  further explored.

Regarding the higher IP CL ratios  [\ion{Si}{x}]/Pa$\beta_{\rm{broad}}$  and [\ion{S}{viii}]/Pa$\beta_{\rm{broad}}$, as discussed in sect. 4, none of the two show a dependance with BH mass.  None of the tested models, included the favoured one  $r =$~3~pc  even account for the loci of the data, with the exception of the $r =$~0.3~pc case  (left panels in both Figs.~\ref{fig:s8pab} and \ref{fig:si10pab}). In the context of the present analysis,  this result is what   is expected  for two reasons: 1), the lack of  a dependence of BH mass with $T_{\rm disc}$  for these CLs  is  attributed to their   IP  falling at the high energy end of the disc spectrum (Fig.~\ref{fig:continuum}),  hence $T_{\rm disc}$ is  not as well sampled as for the  case of lower IP CL for the range of BH mass considered. The lack of dependence with $T_{\rm disc}$ is also indicated by the  CLOUDY predictions,    particularly  for  the [\ion{S}{viii}]/Pa$\beta_{\rm broad}$ case which shows  an almost  straight line along  the temperature axis, as  do the data, whereas  a more erratic behaviour is predicted for  the [\ion{Si}{x}]  case, as also seen the data; 
2) the higher  IP of  [\ion{S}{viii}] and [\ion{Si}{x}] naturally leads to the formation of these ions  at  the innermost edge of the CL region, hence, models with $r$ closer to the central engine should be  favoured, as it appears to be the case. It may also  be considered to decrease the density to foster the production of these ions at larger $r$,  yet,  this  severely penalises the line emissivity, proportional to $ne^2$,    as it can be inferred from  the results (Figs. \ref{fig:s8pab} and \ref{fig:si10pab}).

We finally test the impact of  an accretion rate different from the standard 10\%  Eddington  used in  the models above.  An evaluation of the \LLEdd{} for some of the objects in the sample   points to a range between 1\% and a few 10\% (sect 4.1). 
Narrow Line Seyferts type-I are also predicted to have high Eddington ratios \citep[e.g.][]{kuraszkiewicz_etal2000,panda_etal2019}. Decreasing \LLEdd{} to 1\%  lowers  $T_{\rm disc}$ normalisation in Eq. \ref{eq:tbbb} by factor $\sim$ 1.8,  which is about the same factor reduction introduced by  change of spin from 0.99 to 0.8.
The net effect in the models  would be equivalent to  that produced by the change of spin  shown  in all the figures, Figs.~\ref{fig:si6brg} to~\ref{fig:si10pab}. Effectively, the decrease in $T_{\rm disc}$ implies CL ratios progressively smaller, but the trend of the  models  keeps  similar particularly for cloud distances $r \gtrsim 3$~pc.
Conversely, increasing \LLEdd{} would shift the models in opposite direction towards higher $T_{\rm disc}$. An increase in Eddington ratio by factor 3 yields an increase in $T_{\rm disc}$ by 1.3. Focusing on the  [\ion{Si}{vi}]/Br$\gamma_{\rm broad}$  vs BH mass correlation,    lowering (increasing) the accretion rate or spin just provide  the envelope to  enclose  the scatter  of  the data. The dispersion in the observed  correlation could  largely be    attributed to the range of spin and Eddington accretion rate in the sample.

\begin{figure*}
\includegraphics[width=1.6\columnwidth]{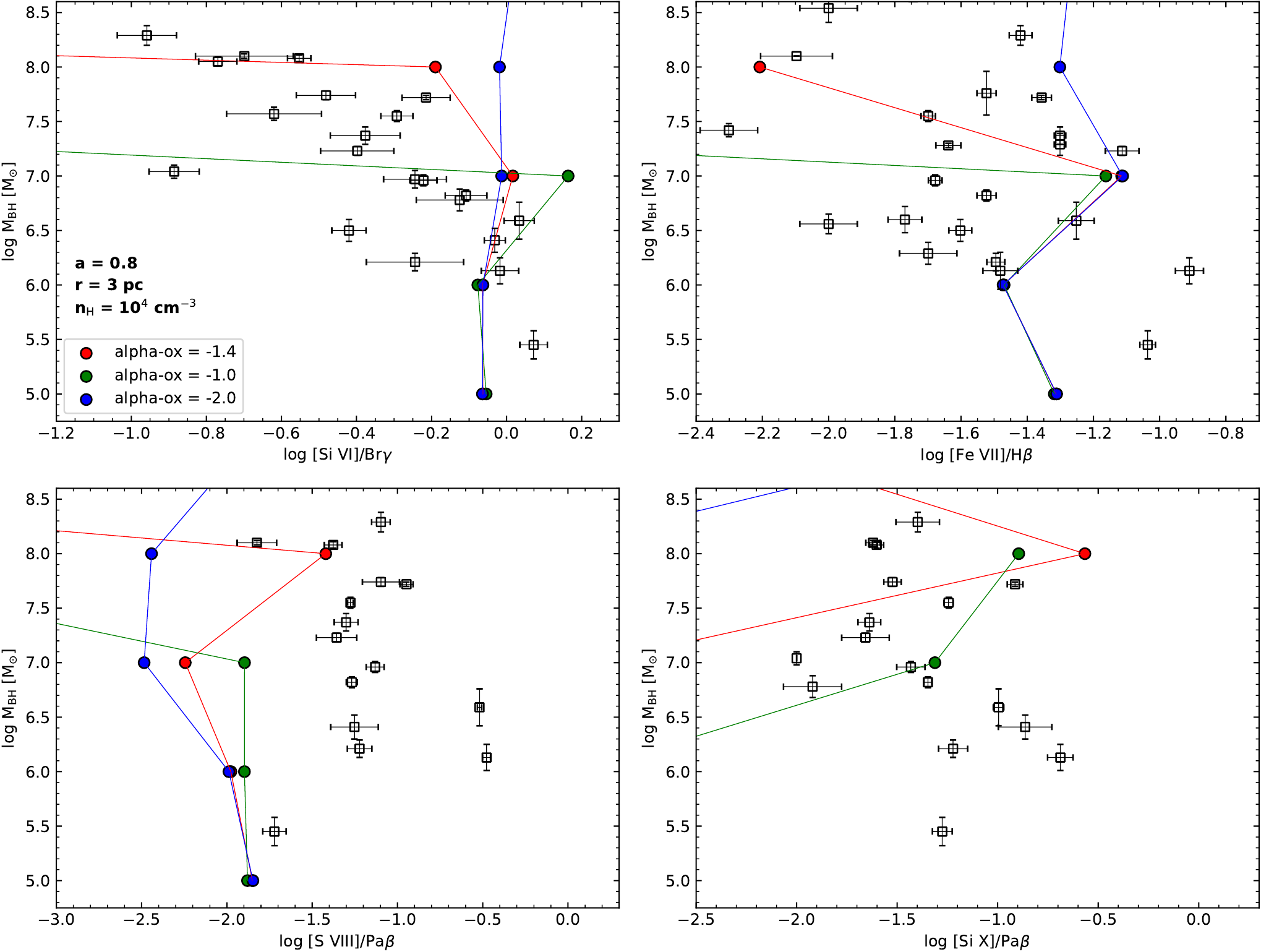}
   \caption{CLOUDY predictions for the four CL ratio vs BH mass diagrams in this work. The ionising continuum is that of eq. 2 but for different values of $\alpha_{\rm ox}$: red refers to  $\alpha_{\rm ox}$= -1.4 - default value used in this work,  sect. 4.2 -   green for $\alpha_{\rm ox}$ = -1,  blue for $\alpha_{\rm ox}$ = -2.   Models are run  for a suit  of parameters that best account for the [Si\,{\sc vi}] - BH mass correlation (Fig.~\ref{fig:si6brg}): density $n_{\rm e} = 10^4$~cm$^{-3}$, distance  $r$ = 3~pc, spin $a$ = 0.8.  All other parameters are as in the default case. Data points as in Fig. \ref{fig:ratios} are black squares.}
    \label{fig:ratios_ss_alphaox}
\end{figure*}

\subsection{Testing the relative-contribution / shape  of the soft X-rays component on coronal emission}
\subsubsection{Modifying $\alpha_{\rm ox}$} 

The IP of the CLs in this work extends  over the soft X-rays, spreading over the 100 -- 350 eV range.  The generic ionising continuum used up to now  (eq. 2, Fig.2), is sampled with a SS disc joined with a power law at the high energies whose relative contribution is fixed with the 
 $\alpha_{ox}$ parameter set to -1.4. Obviously  the contribution of  the hard energy spectrum, which is modulated by  this parameter, affects in different degrees  the CL production, particularly for the highest IP lines. This effect is evaluated bellow.
 
$\alpha_{\rm ox}$ in AGN falls  in the  -1.0  to -2.0 range \citep{avni1986,wilkes1994,lusso2017}, with a typical  value about -1.4 (Zamorani et al. 1981; Ferland et al. 2013).  A new set of models are evaluated  using the generic ionizing spectrum in equ. 2  normalised to either  $\alpha_{\rm ox}$ = -1 or -2. 
To make the comparison with previous results simple, a reduced  set of parameters   that best account for both the  [Si\,{\sc vi}] ratio - BH mass correlation and the data loci in the [Fe VII] ratio vs BH mass are tested, namely   $r = 3$ pc, spin = 0.8 and density = $10^4 cm^{-3}$ (sect. 4.2).  
The results  are in  Figure \ref{fig:ratios_ss_alphaox}, along with      the default   $\alpha_{\rm ox}$ = -1.4 model case in Figs. 3 to 5.   
It can be seen that the $\alpha_{\rm ox}$ = -1.4 case still provides the best envelope of the  [Si\,{\sc vi}] CL ratio vs BH mass correlation - note that one single density is shown.  The new models miss the trend of the data particularly for BH mass above $10^{6.5} Mo$. No  major impact  is sren for  the [Fe\,{\sc vii}] ratio vs BH mass  case, nor for  the higher IP lines  as compared with the default case.
Summarising, making harder / softer  the contribution of the high energy spectrum in the ionising spectrum does not lead to    a better or different account of the data as compared  with the compromise case,   $\alpha_{\rm ox}$ = -1.4.

 \begin{figure}
	\includegraphics[width=\columnwidth]{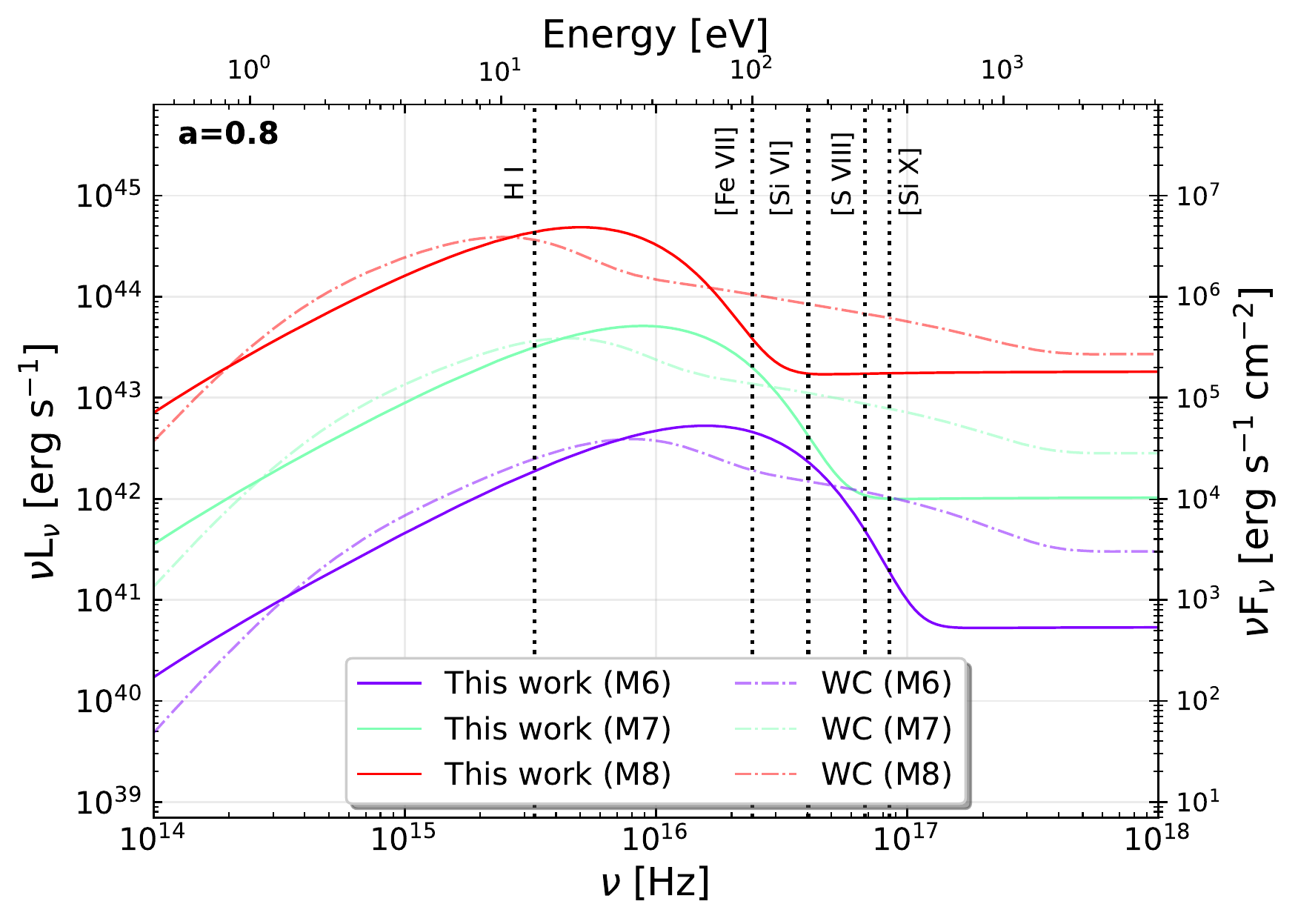}
     \caption{Generic AGN ionizing  continuum as per  Equ. 2  with the addition of   a Warm Comptonizing component (WC in dashed line, following    prescription by  \citet{kubota_done2018}, sect 4.3.2).  The generic case for spin 0.8 (Fig. 2) is shown  for comparison in solid line (note that the spin 0.8 case in Fig. 2 is shown instead with a dots line).  
As in Fig. 2, each case is shown for three BH masses, but one spin, a = 0.8 
     All curves are normalised to \LLEdd{}=0.1.  Vertical lines mark the IPs of the lines used in the analysis.}
     \label{fig:sed_ss_wc}
\end{figure}

\begin{figure*}
	\includegraphics[width=1.6\columnwidth]{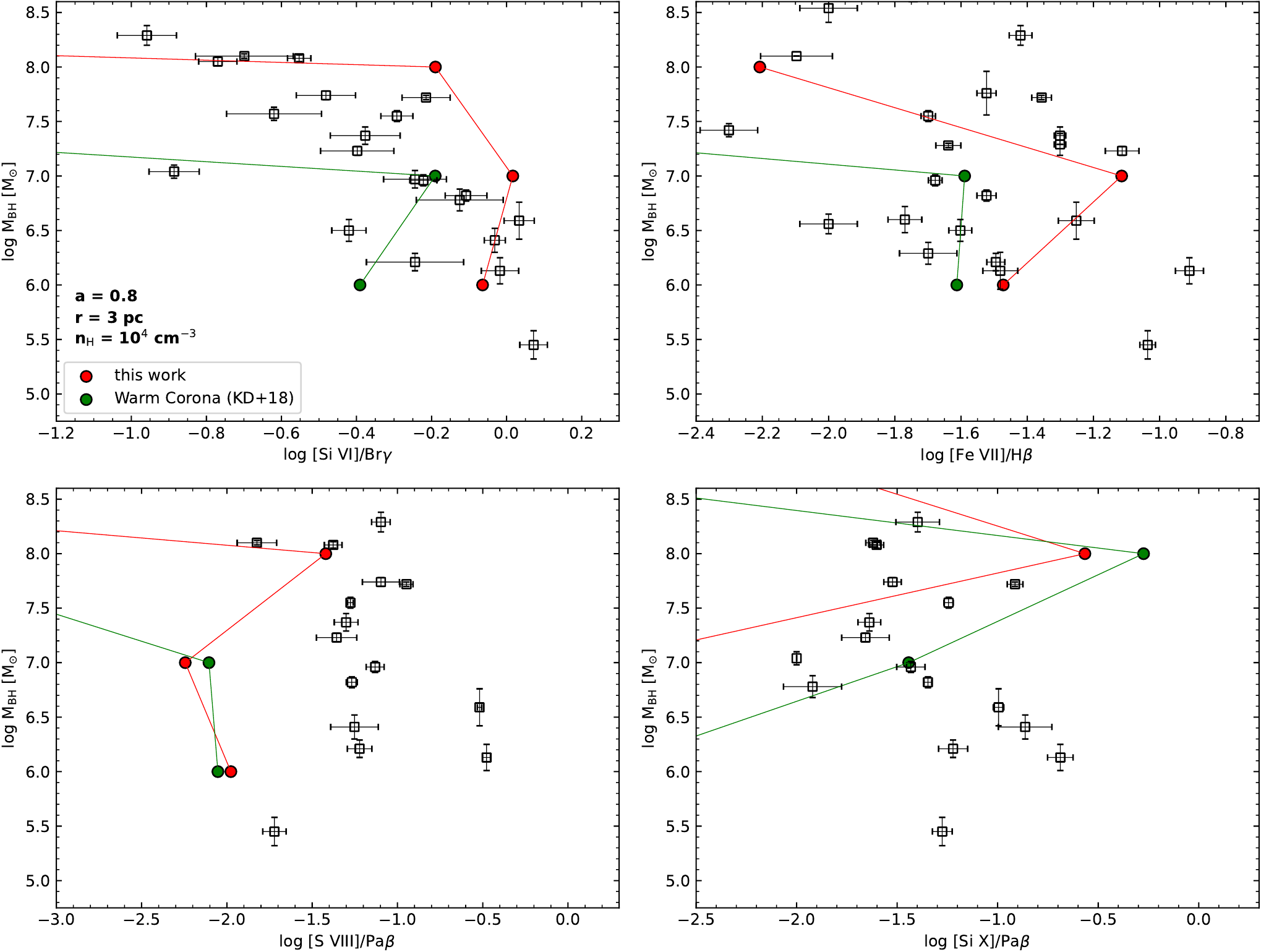}
      \caption{CLOUDY predictions  for the four CL ratio vs BH mass diagrams in this work. In green is shown the case for the generic ionizing continumm with the addition of a warm X-ray component (sect. 4.3.2); in red is  the  default generic  case.  Model parameters are as in sect 4.3.1 case: ne = $10^4 cm^{-3}$,  $r$=3 pc, spin a = 0.8,  $\alpha_{\rm ox}$ = -1.4 with rest of parameters as in sect. 4.1. Data points as in Fig. \ref{fig:ratios} are black squares.}
    \label{fig:sed_ss_wc_ratios}
\end{figure*}

\subsubsection{Adding a warm  Comptonisation component}

A soft X-rays  excess  below $\sim$1keV on top of the nominal high energy power-law  spectrum  is an often  feature in  AGN spectra,   with NLS1 being the most clear representatives \cite[c.f.][]{boller+1996}. Its origin   has been interpreted as  reprocessed emission  from the hot disc corona   or as an additional  warm Comptonising component  dominating the soft X-rays  \cite[e.g.][and references therein]{Fabian+2013,kubota_done2018}.

For the purpose of this work,  to assess the effect of an additional soft X-ray excess component  in  the generic AGN ionizing continuum in this work - Eq.2,  the warm corona component as described in  \citet{kubota_done2018} is taken as a reference and incorporated   as additional component  to the generic continuum. 
The warm corona models are extracted from {\sc agnsed} model \citep{kubota_done2018} using {\sc xspec} \citep{arnaud1996}. The new  ionizing continuum is shown in Fig. \ref{fig:sed_ss_wc} along with the generic one, Eq.~2,  used in this work. The new model is calculated for  Eddington accretion rate of 10 \%, as  the generic case,  but  only the spectrum for  spin = 0.8 is shown. In comparing both spectrum, it can be seen in the figure that with the addition of the warm component the peak of the  ionizing continuum  moves toward cooler temperatures. And because of the new shape of the spectrum,    the IP of the  CLs, particularly of  the reference  ones, [Si\,{\sc vi}] and [Fe\,{\sc vii}~],  fall in a region of the spectrum rather flat with little change  with BH mass. 

Figure \ref{fig:sed_ss_wc_ratios} shows the new model results   for the four CL ratio vs BH mass relations tested in this work.  As in sect. 4.3.1,  models are run for a restricted parameter range, i.e.  ne = $10^4 cm^{-3}$,   $r$=3 pc, and spin a = 0.8, with all other parameters    as in the generic case (sect. 4.1), included  $\alpha_{\rm ox}$ set to -1.4.

The new models provide rather similar  behaviour in   CL ratio vs BH
mass as those produced  with the change of  $\alpha_{\rm ox}$. The
inclusion of the warm soft component predicts  lower values for the
[Si\,{\sc vi}]- and [Fe\,{\sc vii}]-  ratios, as compared with the
effect of changing $\alpha_{\rm ox}$, but otherwise the generic
ionising continuum still provides  the best envelope of the observed
[Si\,{\sc vi}] ratio vs BH mass relation. 

Not major impact is found for the [Fe\,{\sc vii}] vs BH mass diagram, if any  the BH mass range is slightly better covered by the generic model. 
Results are very similar among all  models  regarding   the high IP CL ratios vs BH mass diagrams. Either the range of CL ratios or of BH masses are poorly covered in all cases. There are two aspects to emphasise here. 
1) as discussed in sect. 4.2 the  high IP CL ratios  show no dependance with BH mass, nor any of the generic models hinted for a  dependance either. It was argued as possible  cause the  high IP of the lines which are barely sampled by the  SS disc for the range of BH mass considered. It is not surprising finding similar result  with the new soft X-ray enhanced ionising continuum because of the flattening of  spectrum at the relevant energies region for all the  BH masses.
2) the new models are run for CL cloud distances r = 3~pc whereas smaller r  cover best the bulk of the data as found with the generic ionizing continuum (best coverage is found for r = 0.3 pc, sect. 4.2). Yet, because of  1)  models  for r = 0.3~pc will not provide a distinct result from that obtained with the generic case.

\section{Overall view:  A coronal gas -  BH mass calibration}
\label{sec:remarks}

Using \textit{bona-fide} BH mass estimate from reverberation mapping and the line ratio [\ion{Si}{vi}]~1.963$\mu$m/Br$\gamma_{\rm broad}$ as a genuine tracer of the AGN ionising continuum, a BH-mass scaling relation over almost three orders of magnitude in BH mass, $10^6 - 10^8$ M$_{\odot}$, is found (Fig. \ref{fig:ratios}). The dependence follows a linear regression in log scale $M_{BH} \propto$ ([\ion{Si}{vi}]/Br$\gamma_{\rm{broad}}$)$^{1.99 \pm 0.37}$, with a dispersion in BH mass of 0.47 dex (Sec. \ref{sec3}). Following on the thin accretion disc approximation and after surveying a basic parameter space for coronal gas  production, we believe one of the key parameters driving this correlation is the effective temperature of the accretion disc, the correlation 
 being  formally  in line with the thin disc prediction  $T_{\rm{disc}} \propto {M_{\rm BH}}^{-1/4}$.

On these bases, on  the assumption of a thin disc as the dominant component of the  ionizing continuum  (Fig. 2),  and a suitable range of densities $n_{\rm e} ~ \sim 10^{4 - 6}$~cm$^{-3}$ and  cloud distances, $0.3 < r < 30$ pc,  for CL survival,   
 photoionisation models 
provide a fair representation of  the $M_{\rm BH}$ and [\ion{Si}{vi}]/Br$\gamma_{\rm broad}$ correlation.
 
 No correlation is recovered when normalising the CL emission to narrow \ion{H}{i} gas. This may be due to  two facts: 1) the  much larger emission region covered by  \ion{H}{i}  as compared with that from CL gas; 2)  \ion{H}{i} may be subjected to additional ionisation sources other that the AGN which is not the case  for CL gas.  However, the normalisation to broad HI introduces a complication when comparing with photoionisation models, these been   evaluated for narrow \ion{H}{i} gas. To account for this difference,
a  scaling factor between broad-measured and narrow-predicted \ion{H}{i} gas   is introduced. This scaling factor should be understood as a way to account for the different volume emissivity and density  that characterise     the Broad- and the Coronal- line region.
This scaling   factor was found  in a interactive form, by   shifting  the  models predictions for [\ion{Si}{vi}]/Br$\gamma_{\rm}$ on top of the observed ratios normalised to $Br\gamma_{\rm broad}$.  A factor around 15, unique for all models and objects in the sample, is  found as the best compromise for this CL ratio.   All   other CL -  \ion{H}{i}-broad  ratios used   in this work  were  consequently scaled down from that factor following recombination values.

 No  correlation involving the higher IP lines, [\ion{Si}{x}] and [\ion{S}{viii}], IP $>$250
eV is seen (Fig. 1). We believe this  a natural consequence of the Tdisc  -- BH mass dependence.   Higher IP lines  sample hotter discs, as compared with the [Si VI] IP, and in turn smaller BH masses, not covered by this sample, hence the absence of a positive trend  for the range of masses in this work. It follows that  a lack of dependence with $T_{\rm disc}$ should then be expected. This is    consistent with the photoionisation
predictions  as inferred from  the same set of models used for the [Si VI] case (Figs. 5 and 6).
The high IP CL are however expected to show a dependence with  BH masses below 10$^6$ M$_{\odot}$ (Fig. 2, Eq. 3),  making them suitable BH mass scaling indicators for intermediate BH masses, as suggested by \citep{cann+18}. Testing this low mass end is currently limited due to the  unavailability of accurate BH masses and suitable CL data.

Thus, the  BH mass - CL dependence  appears sensitive to the ionisation potential of the CL employed. The use of [\ion{Si}{vi}]~1.963~$\mu$m, IP = 167 eV, restricts the dependence to BH masses in the range $10^6 - 10^8$~M$_{\odot}$, presumably because of the disc temperature, for these masses, favours the production of $Si^{+5}$ best (Fig. \ref{fig:continuum}). In the same line of reasoning, an equivalent dependance involving   other CL with IP close  to the disc peak emission  for these range of BH mass should  be expected. Such is the case of  e.g. [\ion{Fe}{vii}]~$\lambda$6087\AA, IP =100 eV, but a scatter relation is found instead. Photoionisation predictions do not   hint   for a  dependance with Temperature either (Fig. 4). Thus  no clear conclusion for the potential of this line as gas temperature can be assessed.

Other suitable CL are those of   Ne$^{4+}$ IP = 97 eV. Yet, the UV lines are much subjected to reddening,  those in the mid-IR are optimal but available for  few sources with accurate BH mass determination. The analysis of these few sources shows indeed a trend with BH mass but the statistic is  insufficient to establish a correlation, the analysis is in progress.

Above $10^8$ M$_{\odot}$ BH mass,  disc temperatures are foreseen in the $10^5$~K regime in the disc approximation.  Lower ionisation lines would then be more favoured than the higher ones. 
The [\ion{Si}{vi}]/Br$\gamma_{\rm broad}$ -BH mass correlation points in that direction, with  [\ion{Si}{vi}]/Br$\gamma_{\rm broad}$ decreasing  with increasing BH mass (Fig. \ref{fig:ratios}). The high BH mass tail in the local universe is the realm of elliptical and bulge dominated objects often associated with LINER (Low Ionisation Nuclear Emitting Regions) activity.   The parsec-scale spectral energy distribution   of some of these sources strictly limit the temperature of the  disc  to  $ < 10^5$ K  (Fernandez-Ontiveros et al. submitted).
Parsec-scale  near-IR observations have so far proven the elusiveness of CL emission in a few of these cases \citep{muller-sanchez+13,mazzalay_etal2014}, in line with that prediction. These sources would appear as upper limits at the high mass range in Fig.~\ref{fig:ratios}. 

Because of the different physical properties of the objects in the  sample, some of the scatter in the proposed  BH mass scaling relation should be intrinsic. In the proposed scenario,  prime properties that should differ between objects are the  accretion rate and spin.  They affect the observed  CL ratios dependence with BH mass in different degrees. Best agreement with  the observed [\ion{Si}{vi}]~1.963$\mu$m/Br$\gamma_{\rm broad}$ -- BH mass  trend is achieved for spins above 80\% and Eddington accretion rate of 10\% (Fig. 3).  A increase  in  spin,  or decrease  in accretion rate,  translates into a similar effect, namely  a progressive shift to higher, or lower, $T_{\rm disc}$ for  same BH mass, by factor of about 1.8 from spin =  0.8  to 0.99 or Eddington accretion rate from 10\% to 1\% (Fig. 3). Higher  Eddington rates  up to 30\%,  in line with some objects in the sample, would lead to a slight increase in  $T_{\rm disc}$ by 1.3. In practise, the decrease in $T_{\rm disc}$ makes [\ion{Si}{vi}]/Br$\gamma$ progressively less sensitive to lower  BH mass as the peak temperature of the disc moves away from the optimal energy for producing $Si^{+5}$ whereas the increase in $T_{\rm disc}$ by similar factor would favour the diagnosis for lower BH masses. 

Finally,   changes on the relative contribution of a soft X-ray excess component on the CL ratio - BH mass dependences     was evaluated.  This was done by  either  changing  the  $\alpha_{ox}$ parameter or adding  a warm X-ray corona component  in the standard ionising continuum used in this work.  None of these variants    provide a significative  account  of the BH mass - [Si VI] ratio correlation (Fig. 7, 9).  Nor, a  significant result    was found for the higher IP CLs, this even though   the modification of the ionising spectrum at the relevant energies to produce these lines. An important  result from the photoionisation modelling  is that  to produce the range of  CL ratios observed at these  high IPs, clouds should be  less than few tenth of parsec  from the centre (Figs. 5, 6). Still,  to verify the potential of these high IP lines as BH mass scale tracers, the lower BH mass range has to be tested. The detection of coronal [Fe X],  IP = 240 eV, in a large sample of dwarf galaxies by \cite{Molina+2021} opens an exciting avenue.  
 
With a final compendium of 21 objects, the dispersion in BH mass in the proposed  calibration is 0.47 dex (1$\sigma$). In comparison,  a dispersion of  0.44 dex is inferred  from the $M-\sigma$ relation in 49 galactic bulges with direct dynamical BH mass estimate \citep{gultekin+09}. The   intrinsic scatter in the mass - luminosity relations is  in the 40\% range \citep{kaspi_etal2005}, mostly driven by differences in optical - UV continuum shape. In the present [Si VI] - BH mass correlation, variations in the ionising continuum shape  e.g. the soft X-ray excess slope and contribution, should contribute to the  scatter but we consider that  a minor effect  as the production of the CL gas 
is mostly driven by the peak energy in the UV.

The present BH mass scaling relation is restricted to Type~1 AGN including narrow line Type~I. The limitation is driven by the imposition of including  {\it bona-fide} BH masses only, and the need to normalise to broad \ion{H}{i} gas.  We are nonetheless examining possibilities to extend it to Type 2.  Source variability or changes in instrumental setup are not an issue. We found that most of the scatter is chiefly driven by  differences in BH spin and accretion rate. The new  scaling offers an economic,  and physically motivated alternative for BH estimate using single epoch spectra, avoiding  large telescope time (reverberation mapping) or absolute flux calibration (the continuum luminosity method). With James Webb Space Telescope and big surveys in the IR region, large samples of AGNs could be weighted using this approach.
 
\section*{Acknowledgements}

We are grateful to the referee, Hermine Landt, for her positive suggestions to improve this manuscript. We are grateful to H. Netzer,  E. Churazov, B. Czerny, A. Askar for discussions.
MAP thanks the support  of the Excellence Cluster ORIGINS which is funded by the Deutsche Forschungsgemeinschaft (DFG, German Research Foundation) under Germany Excellence Strategy - EXC-2094 - 390783311. A.R.A acknowledges partial support from CNPq Fellowship (312036/2019-1	 \& 203746/2017-1). S.P acknowledges partial support from the Polish Funding Agency National Science Centre, project 2017/26/\-A/ST9/\-00756 (MAESTRO  9) and the computational facility at Nicolaus Copernicus Astronomical Centre.

\section*{Data Availability}

Data products will be shared on reasonable request to the corresponding author.



\bibliographystyle{mnras}
\bibliography{article} 




\appendix

\section{Optical and NIR Spectra}


 \begin{figure*}
	\includegraphics[scale=0.8]{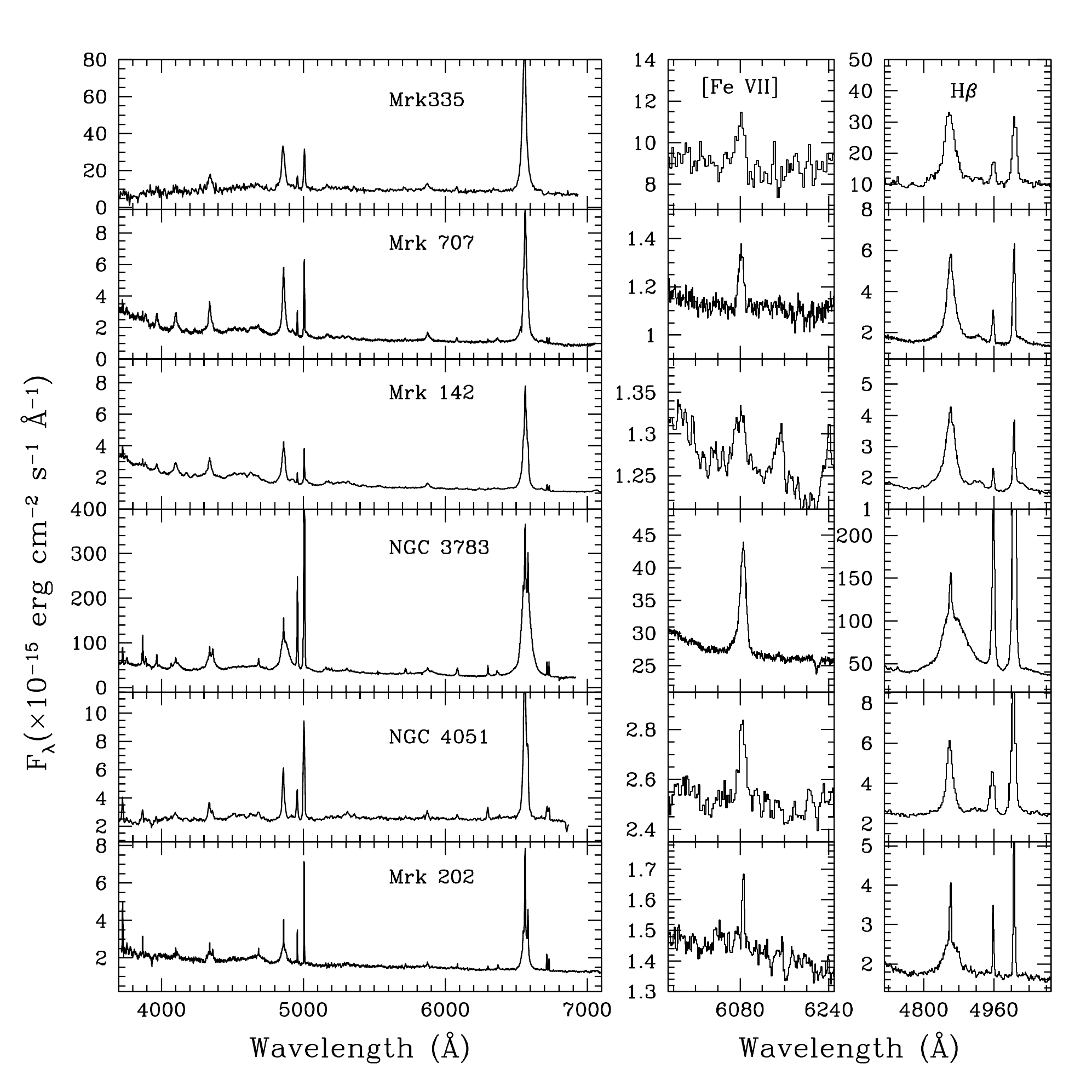}
    \caption{Optical Spectra of the AGN sample in rest wavelength. For each galaxy, the left panel shows the full spectrum. The two smaller panels to the left shows [\ion{Fe}{vii}]~$\lambda$6087 and the H$\beta$ line. }
    \label{fig:optspec1}
\end{figure*}

 \begin{figure*}
	\includegraphics[scale=0.8]{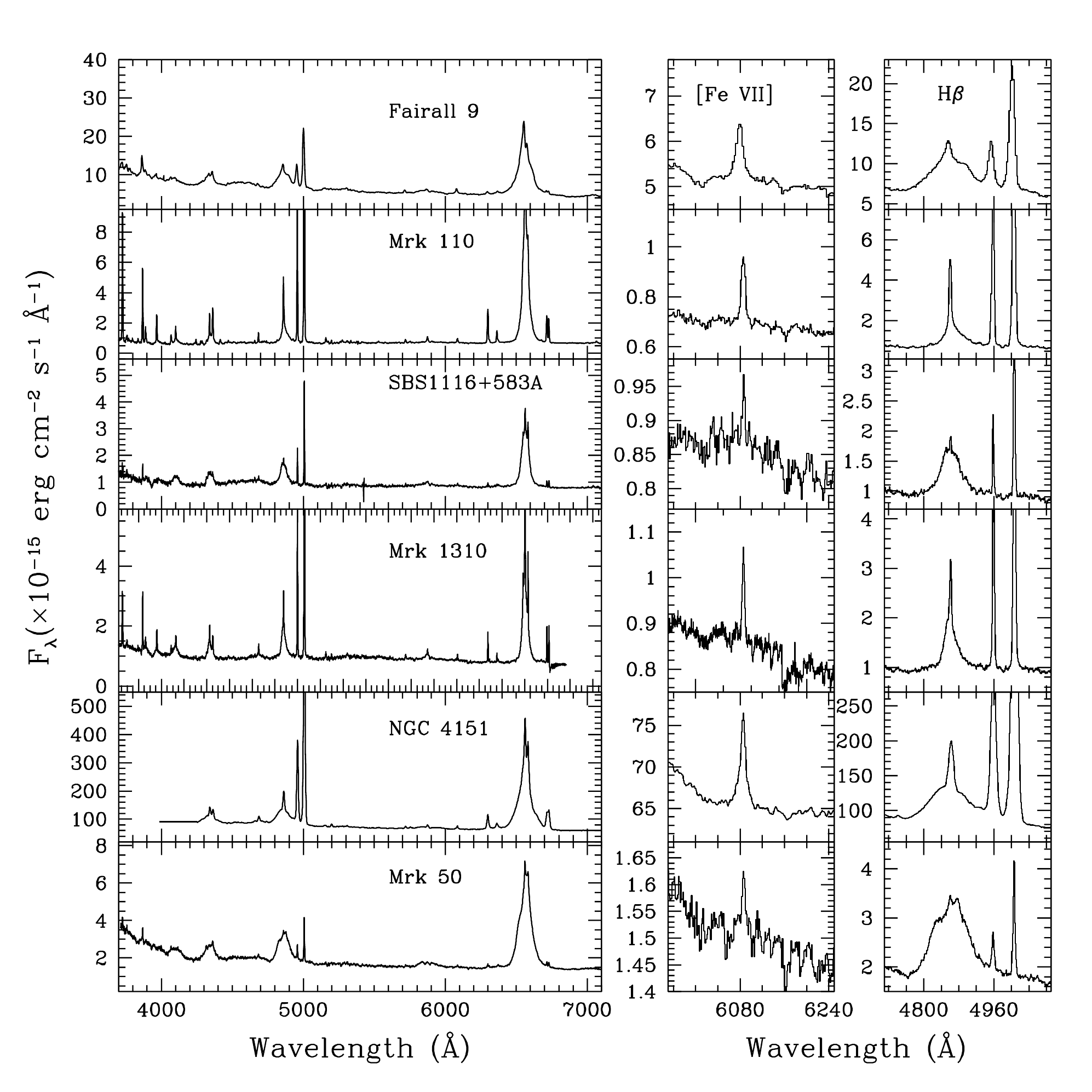}
    \caption{Cont. Fig.~\ref{fig:optspec1}.}
    \label{fig:optspec2}
\end{figure*}

\begin{figure*}
	\includegraphics[scale=0.8]{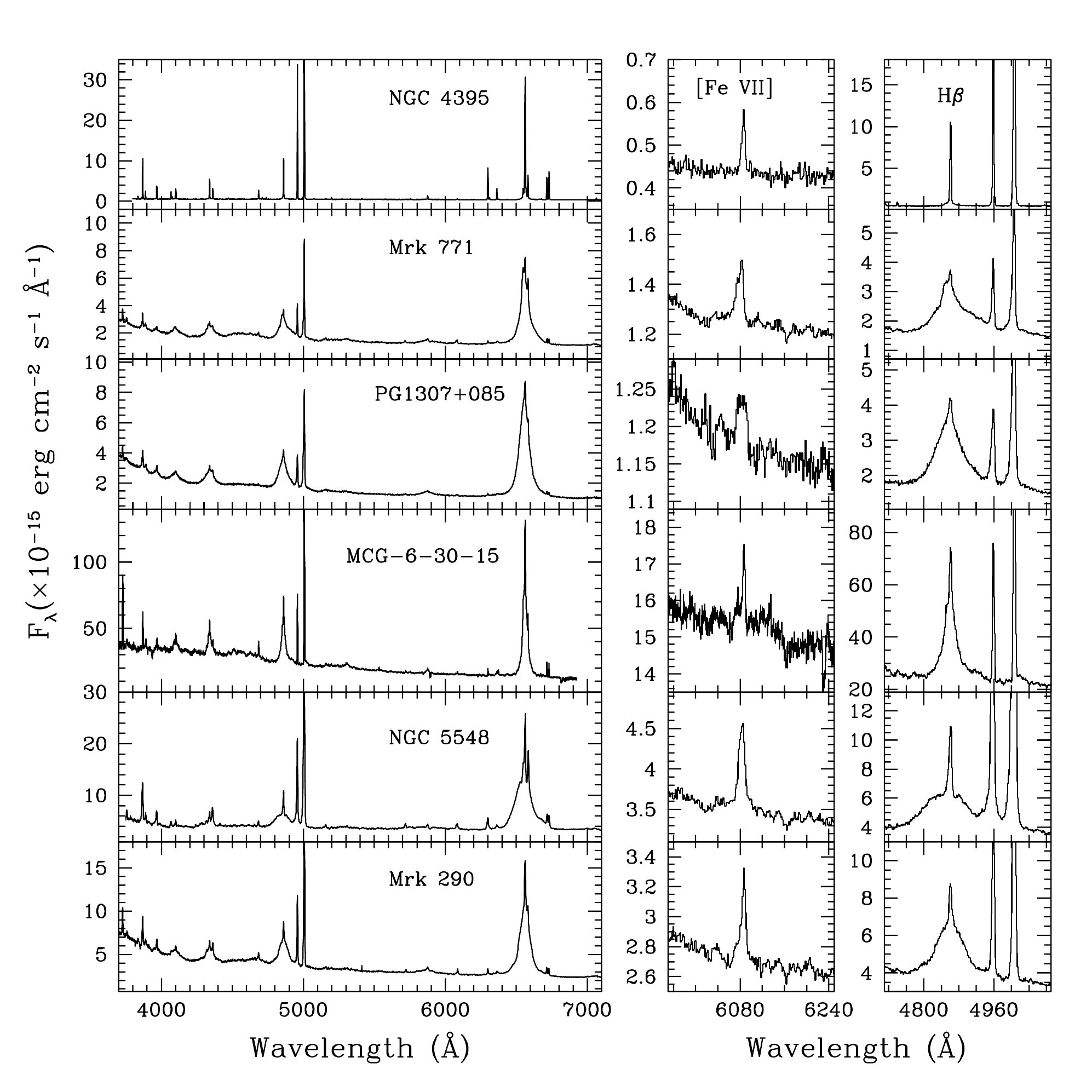}
    \caption{Cont. Fig.~\ref{fig:optspec1}. }
    \label{fig:optspec3}
\end{figure*}

\begin{figure*}
	\includegraphics[scale=0.8]{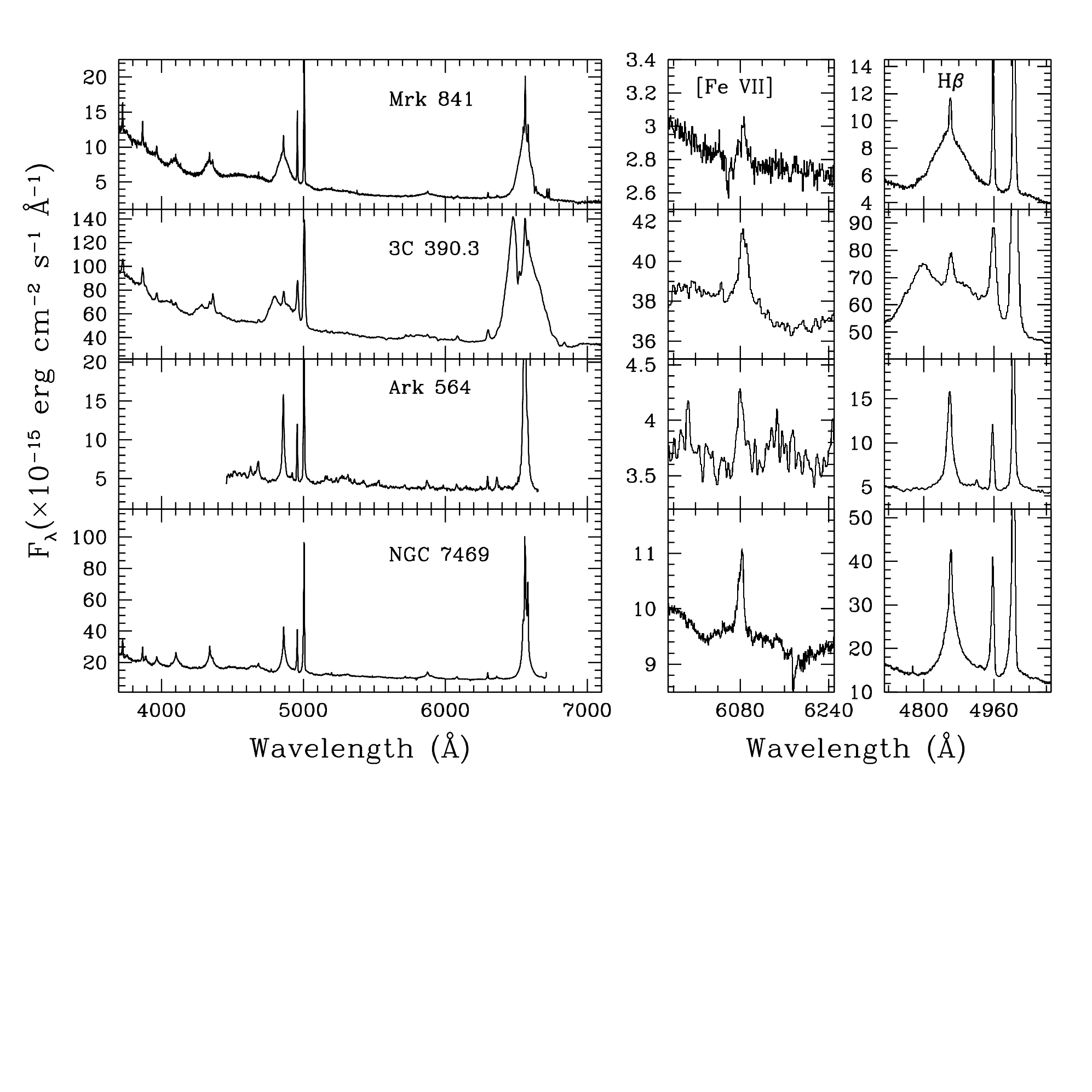}
    \caption{Cont. Fig.~\ref{fig:optspec1}. }
    \label{fig:optspec4}
\end{figure*}

\begin{figure*}
	\includegraphics[scale=0.9]{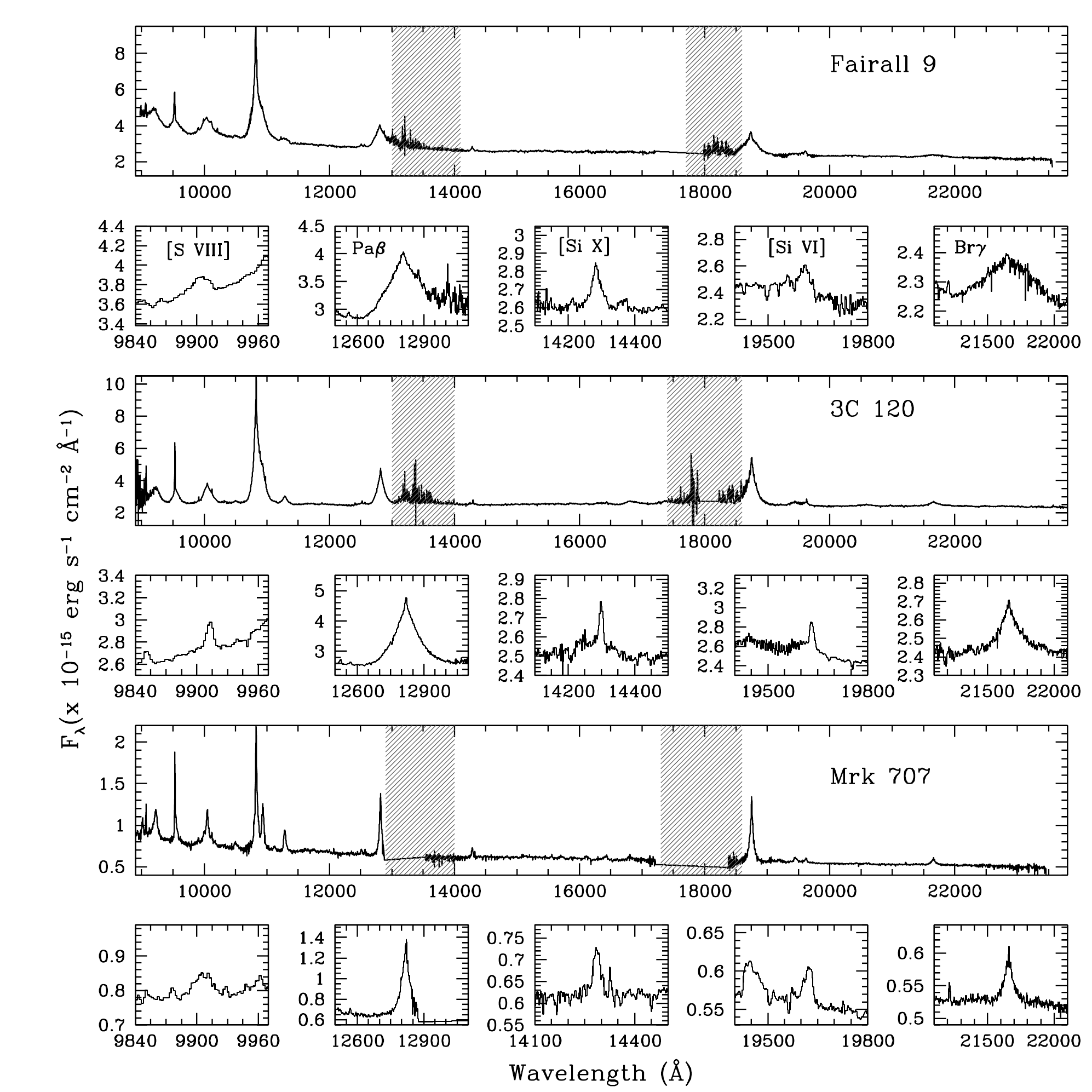}
    \caption{NIR spectra of Fairall~9 (top panels), 3C~120 (middle panels) and Mrk~707 (bottom panels) in rest wavelength.  For each AGN, the larger panel displays the observed spectrum in the 9000$-$23000~\AA\ range. The smaller panels in the following row show a zoom around the most relevant lines to this work. The shaded areas mark regions of bad atmospheric transmission.}
    \label{fig:nirspec1}
\end{figure*}

\begin{figure*}
	\includegraphics[scale=0.9]{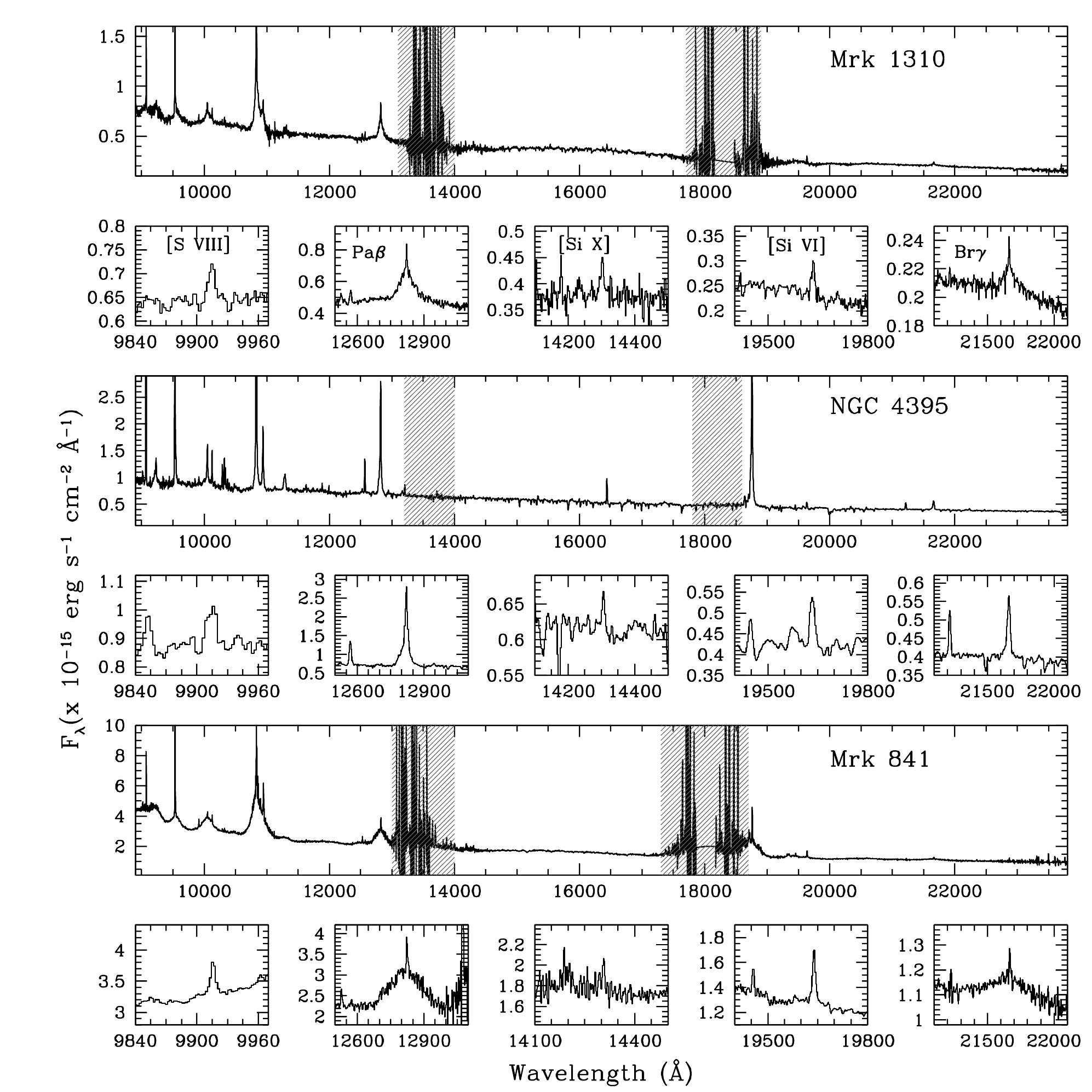}
    \caption{Same as Figure~\ref{fig:nirspec1} for Mrk~1310 (top panels), NGC~4395 (middle panels) and Mrk~841 (bottom panels).}
    \label{fig:nirspec2}
\end{figure*}

\begin{figure*}
	\includegraphics[scale=0.9]{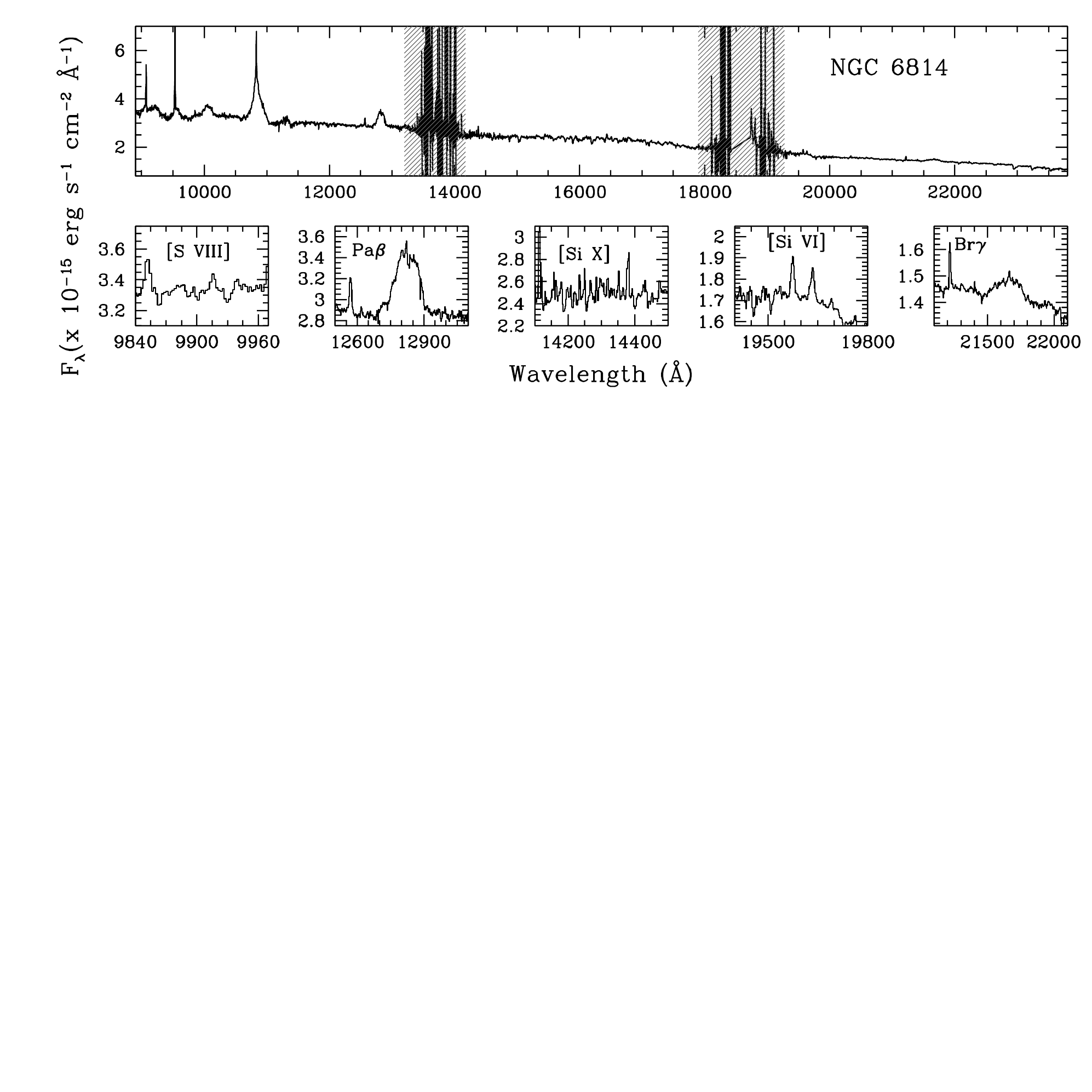}
    \caption{Same as Figure~\ref{fig:nirspec1} for NGC~6814.}
    \label{fig:nirspec3}
\end{figure*}

\section{Emission line fluxes measured for the galaxy sample}

\begin{table*} 
	\caption{Measured broad line H\,{\sc i} and coronal line fluxes, in units of 10$^{-15}$~erg~cm$^{-2}$~s$^{-1}$, for the galaxy sample.}
	\label{tab:fluxes}
	\begin{tabular}{lccccccc}
	\hline
Galaxy & H$\beta$ & Pa$\beta$ & Br$\gamma$ & [Fe\,{\sc vii}] & [S\,{\sc viii}] & [Si\,{\sc x}] & [Si\,{\sc vi}] \\
       & $\lambda$4861 & $\lambda$12818 & $\lambda$21657 & $\lambda$6087 & $\lambda$9913 & $\lambda$14203 & $\lambda$19630 \\ 
        \hline
Mrk\,335 &  712.68$\pm$44.13 & 170$\pm$5 & 26.7$\pm$3.1 & 55.12$\pm$5.50 & 3.76$\pm$1.00 & 7.45$\pm$2.00 & 10.6$\pm$2.0 \\
Fairall~9 & 490.25$\pm$30.12 & 107.95$\pm$8.40 & 43.1$\pm$4.7 & 18.4$\pm$1.33 & 4.23$\pm$0.52 & 8.16$\pm$0.67 & 4.00$\pm$0.52 \\
NGC\,863 &  ... & 49$\pm$3 & 10.40$\pm$2.94 & ... & ... & ... & 2.54$\pm$0.14 \\
3C\,120   & ... & 222.9$\pm$5.94 & 46.34$\pm$6.55 & ... & 2.97$\pm$0.17 & 9.96$\pm$2.47 & 15.48$\pm$1.96 \\
Mrk\,707 & 107.67$\pm$3.31 & ... & 4.76$\pm$0.39 & 2.60$\pm$0.23 & 1.4$\pm$0.27 & 3.18$\pm$0.31 & 1.82$\pm$0.09 \\
Mrk\,110  & 52.22$\pm$1.43 & ... & ... & 2.43$\pm$0.08 & ... & ... & ...  \\
NGC\,3227 &  ... & 168.74$\pm$10.23 & 20.0$\pm$3.92 & ... & 1.97$\pm$0.6 & ... & 14.9$\pm$2.6 \\
Mrk\,142 & 86.05$\pm$0.70 & ... & ... & 1.67$\pm$0.35 & ... & ... & ... \\
SBS\,1116+583A	& 47.23$\pm$1.93 & ... & ... & 0.46$\pm$0.08 & ... & ... & ... \\
PG\,1126-041 & ... & 101.83$\pm$3.7 & 13.3$\pm$0.73 & ... & 2.58$\pm$0.21 & 4.32$\pm$0.45 & 3.71$\pm$0.17 \\ 
NGC\,3783 & 4693.0$\pm$157.0 & 348.70$\pm$19.06 & 60.40$\pm$9.82 & 242.0$\pm$6.48 & 8.12$\pm$0.86 & 17.51$\pm$2.67 & 25.6$\pm$3.71 \\
Mrk\,1310 & 34.9$\pm$0.72 & 11.4$\pm$1.1 & 1.4$\pm$0.4 & 1.10$\pm$0.05 & 0.79$\pm$0.10 & 0.83$\pm$0.19 & 0.74$\pm$0.11 \\ 
NGC\,4051 &  48.22$\pm$2.44 & 66.6$\pm$1.7 & 13.1$\pm$0.8 & 5.95$\pm$0.51 & 13.7$\pm$2.0 & 22.2$\pm$1.1 & 12.5$\pm$1.2 \\
NGC\,4151 &  7585$\pm$79 & 712.48$\pm$8.77 & 125.0$\pm$10.81 & 151.0$\pm$3.47 & 40.5$\pm$2.1 & 37.7$\pm$1.6 & 64.0$\pm$1.3 \\
Mrk\,202 & 41.9$\pm$2.14 & ... & ... & 1.41$\pm$0.15 & ... & ... & ... \\
Mrk\,766 & 829$\pm$30 & 117.8$\pm$1.8 & 20.0$\pm$2.26 & 24.0$\pm$1.20 & 5.3$\pm$0.2 & 6.3$\pm$0.4 & 15.6$\pm$0.9 \\
Mrk\,50 & 160$\pm$5.0 & ... & ... & 0.77$\pm$0.15 & ... & ... & ... \\
NGC\,4395 & 11.0$\pm$0.2 & 30.5$\pm$1.1 & 1.9$\pm$0.2 & 1.03$\pm$0.06 & 1.6$\pm$0.2 & 0.57$\pm$0.09 & 2.2$\pm$0.1 \\
Mrk\,771 & 125.25$\pm$2.44 & ... & ... & 3.88$\pm$0.23 & ... & ... & ... \\
NGC\,4748 &  ... & 60.1$\pm$2.2 & 9.7$\pm$0.5 & ... & 8.2$\pm$2.5 & 3.4$\pm$1.1 & 9.0$\pm$0.3 \\
PG\,1307+085 & 172.0$\pm$2.4 & ... & ... & 1.65$\pm$0.28 & ... & ... & ... \\
MGC-6-30-15 &	1100.0$\pm$4.0 & ... & ... & 18.8$\pm$2.15 & ... & ... & ... \\
NGC\,5548 & 314.0$\pm$9.77 & 49.3$\pm$2.9 & 16.3$\pm$2.0 & 13.9$\pm$0.65 & 6.0$\pm$0.4 & 5.6$\pm$0.4 & 10.0$\pm$0.9 \\
PG1448+273 &  ... & ...  & 2.40$\pm$0.10 & ... & 0.66$\pm$0.06 & ... & 1.6$\pm$0.2 \\
Mrk\,290 & 239.0$\pm$5.25 & ... & ... & 5.44$\pm$0.36 & ... & ... & ... \\
Mrk\,841 & 435.56$\pm$12.0 & 146.7$\pm$10.4 & 26.1$\pm$7.5 & 3.42$\pm$0.94 & 3.5$\pm$0.2 & 2.2$\pm$0.5 & 5.2$\pm$0.1 \\
3C\,390.3  & 433.12$\pm$7.1 & ... & ... & 7.82$\pm$0.39 & ... & ... & ... \\
NGC\,6814 & ... & 81.0$\pm$13.3  & 11.9$\pm$3.5 & ... & 0.55$\pm$0.23 & ... & 2.00$\pm$0.29 \\
Mrk\,509 &  ... & 1824.7$\pm$77.9 & 349.0$\pm$21.8 & ... & ... & ... & 58.8$\pm$5.1 \\
Ark\,564 & 160.1$\pm$2.2 & 59.0$\pm$1.5 & 5.7$\pm$0.4 & 9.0$\pm$1.14 & 5.9$\pm$0.4 & 17.9$\pm$0.3 & 6.2$\pm$0.3 \\
NGC\,7469 & 854.59$\pm$12.0 & 153.1$\pm$7.3 & 20.7$\pm$1.4 & 17.9$\pm$0.96 & 5.6$\pm$0.8 & 11.4$\pm$1.3 & 12.4$\pm$0.5 \\
\hline
\end{tabular}
\end{table*}


\bsp	
\label{lastpage}
\end{document}